\documentclass[%
 reprint,
 superscriptaddress,
 amsmath,amssymb,
]{revtex4-1}

\usepackage{graphicx}
\usepackage{dcolumn}
\usepackage{bm}
\usepackage{makecell}
\usepackage{epsfig}
\usepackage{subcaption}
\usepackage{verbatim}
\usepackage{setspace}
\usepackage{filecontents}
\usepackage{xr}
\graphicspath{{../figs/}}
\usepackage[mathlines]{lineno}

\begin{document}


\title{Critical percolation threshold restricts late-summer Arctic sea ice melt pond coverage}
\author{Predrag Popovi\'{c}}
 \altaffiliation{All correspondence should be directed to Predrag Popovi\'{c}}
  \email{arpedjo@gmail.com}
  \affiliation{%
 The University of Chicago, The Department of the Geophysical Sciences\\
}%
\author{Mary C. Silber}%
\affiliation{%
The University of Chicago, Department of Statistics and Committee on Computational and Applied Mathematics\\
}%
\author{Dorian S. Abbot}
\affiliation{%
The University of Chicago, The Department of the Geophysical Sciences\\
}%




\date{\today}

\begin{abstract}

During the summer, vast regions of Arctic sea ice are covered by meltwater ponds that significantly lower the ice reflectivity and accelerate melting. Ponds develop over the melt season through an initial rapid growth stage followed by drainage through macroscopic holes. Recent analysis of melt pond photographs indicates that late-summer ponds exist near the critical percolation threshold, a special pond coverage fraction below which the ponds are largely disconnected and above which they are highly connected. Here, we show that the percolation threshold, a statistical property of ice topography, constrains pond evolution due to pond drainage through macroscopic holes. We show that it sets the approximate upper limit and scales pond coverage throughout its evolution after the beginning of drainage. Furthermore, we show that the rescaled pond coverage during drainage is a universal function of a single non-dimensional parameter, $\eta$, roughly interpreted as the number of drainage holes per characteristic area of the surface. This universal curve \linelabel{B6}allows us to formulate an equation for pond coverage time-evolution during and after pond drainage that captures the dependence on environmental parameters and is supported by observations on undeformed first-year ice. \linelabel{B18}This equation reveals that pond coverage is highly sensitive to environmental parameters, suggesting that modeling uncertainties could be reduced by more directly parameterizing the ponds' natural parameter, $\eta$. Our work uncovers previously unrecognized constraints on melt pond physics and places ponds within a broader context of phase transitions and critical phenomena. Therefore, it holds promise for improving ice-albedo parameterizations in large-scale models. 

\end{abstract}

\maketitle

\section{Introduction}

\begin{figure*}
\centering
\includegraphics[width=0.7\linewidth]{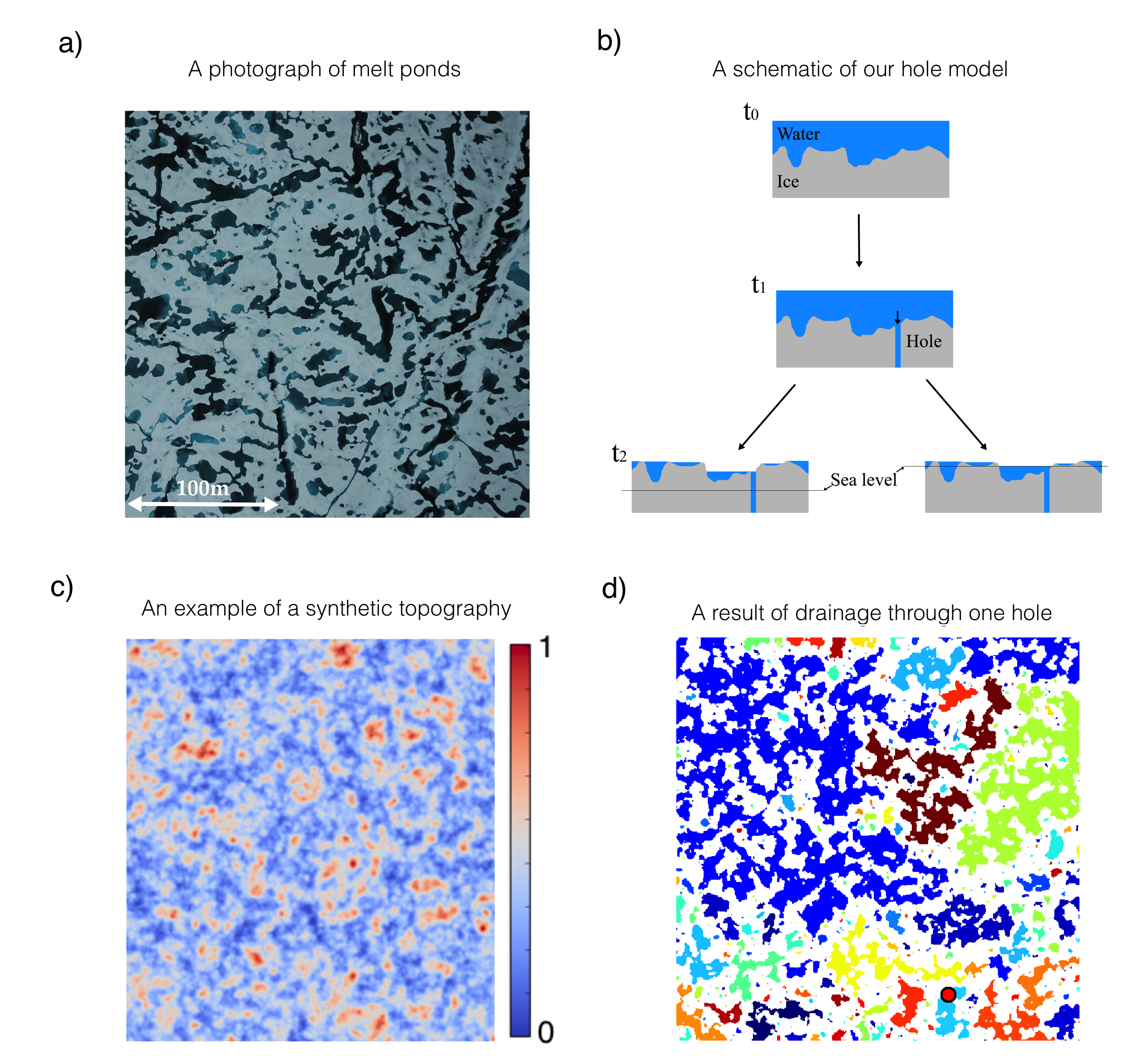}
\caption{ a) A photograph of melt ponds taken during the 2005 HOTRAX mission. b) A sketch of the hole drainage model. The model begins with an ice surface, a large fraction of which is flooded. A hole opens at a random location on the ice and starts draining the ponds that are connected to it. As the drainage progresses, some regions of the ice become disconnected from the hole and can no longer drain through it. Drainage stops when either the hole is exposed to the atmosphere or when the water level of the pond connected to the hole reaches sea level. c) An example of randomly generated ``snow dune'' topography. Red colors stand for topographic highs while blue colors stand for topographic lows. The upper bound on the scale bar, here set to 1, is arbitrary. This panel is taken from \citet{popovic2020snow}. d) Ponds after drainage on the topography in panel c) through the hole marked with a red dot. Different colors correspond to distinct ponds.}
\label{fig:1}
\end{figure*}

\noindent
Arctic sea ice covers a vast area of nearly 15 million square kilometers at its peak annual extent. It sculpts the Arctic environment, supports its ecosystem, and presents a significant obstacle to trade \citep{perovich2009loss}. In recent years, sea ice has been rapidly disappearing at a rate underestimated by most climate models \citep{stroeve2007arctic}. These climate models cannot resolve processes on scales smaller than tens of kilometers, and the disagreement with observations is largely attributed to such unresolved processes \citep{holland1999role}. A notable example of a small-scale process contributing to the uncertainty of the large-scale predictions is meter-sized melt ponds that form on the ice surface during summer (Fig. \ref{fig:1}a). Melt ponds absorb roughly twice as much solar radiation as surrounding bare ice, which significantly accelerates ice melt \citep{perovich1996optical}. Several works have shown the crucial role ponds play in predicting the state of sea ice \citep{schroder2014september,holland2012improved}. For example, \citet{schroder2014september} demonstrated that the summer sea ice minimum extent can be accurately predicted based only on the spring melt pond extent. Currently, the most common tactic for modeling melt ponds is to try to represent as many of the physical processes that contribute to pond evolution as realistically and as comprehensively as possible \citep{luthje2006modeling,flocco2007continuum,skyllingstad2009simulation,skyllingstad2015simulation}. Although such models capture many properties of melt pond evolution, it is unclear how assumptions and details of these complex models might change in a warmer climate. This suggests an opportunity for modeling based on universal properties of sea ice that will likely remain unchanged in a warmer climate.

Melt ponds develop in several stages that depend on the microstructure of the underlying ice \citep{polashenski2012mechanisms,landy2014surface,polashenski2017percolation}. As sea ice forms from freezing salty water, brine pockets and channels fill the ice interior \citep{perovich1996quantitative,cole1998observations,golden2007thermal}. When the summer melt season begins, fresh water from melting snow penetrates into the cold ice interior through these brine structures, freezes, and thereby plugs the pathways connecting the ice surface to the ocean. The lack of pathways to the ocean allows water to collect in ponds at the top of the ice \citep{polashenski2017percolation}. Since the ice is relatively flat, ponds grow rapidly during this time, and sometimes end up covering the majority of the ice surface at their peak. This rapid growth stage is usually known as ``stage I'' of pond evolution. Later, as ice warms, some of the frozen freshwater plugs melt, once again opening the pathways to the ocean and draining the ponds. As relatively warm pond water flows through these channels, it can expand them into large holes that can drain very large areas in a matter of hours \citep{polashenski2012mechanisms}. More such holes open with time, and, over the course of several days, pond coverage falls to its minimum. The period during which ponds that are above sea level drain to reduce their hydraulic head is known as ``stage II'' of pond evolution. Afterwards, the remaining ponds correspond to those regions of ice surface that are below sea level. From this point, pond coverage slowly grows as the ice thins and more of the ice surface falls below sea level. This is known as ``stage III'' of pond evolution. Observations show that during stage II, meltwater is mainly drained through large holes, while during stage III it is mostly drained through microscopic pathways in the ice \citep{polashenski2012mechanisms}. 

Recently, in \citet{popovic2018simple}, we analyzed photographs of melt ponds and showed that the post-drainage melt pond geometry can be accurately described by a purely geometrical model where ponds are represented as voids that surround randomly sized and placed circles, which can be loosely interpreted as snow dunes. Surprisingly, we found that in order to match various statistics derived from images of late-summer ponds, the fraction of the surface covered by voids had to be tuned to a special value: the percolation threshold. The concept of a percolation threshold coverage fraction, $p_c$, was developed in physics and applied to modeling diverse phenomena ranging from electrical transport in disordered media to turbulence \citep{isichenko1992percolation}. In the idealized setting of an infinite plane covered randomly by objects that can overlap to form larger clusters, the percolation threshold is the coverage fraction below which only finite connected clusters can exist and above which an infinite cluster that spans the domain forms. Exactly at $p_c$, connected clusters of all sizes exist, and the system becomes scale-invariant. Close to this threshold, percolation models exhibit \textit{universality} \citep{goldenfeld1992lectures} which means that much of the system behavior does not depend on details such as, for example, the shape of connecting objects. Generally speaking, universality is often observed near the critical point of continuous phase transitions. Thus, for example, universality allows both the magnetic phase transition and the phase transition of fluids near the critical point to be accurately described by the idealized Ising model \citep{goldenfeld1992lectures,nishimori2010elements}. 

Our observation that late-summer ponds seem to be organized close to the percolation threshold presents a puzzle and suggests that there is a mechanism driving the ponds to this threshold. Here, we show that drainage through large holes can account for this observation. Specifically, we develop an idealized model of pond drainage through large holes to show how this mechanism drives melt pond evolution to lie below the percolation threshold. We then use this model to formulate a pond fraction evolution equation that reveals the connections between the pond evolution and measurable physical parameters. \linelabel{B2}Even though our hole drainage model explicitly represents the drainage stage (stage II), we show that it can also be used to understand elements of pond evolution during stage III. By clarifying the relation of pond coverage to the percolation threshold, we place melt pond formation within the broader context of critical phenomena and phase transitions.

\linelabel{A4}Our approach to modeling melt ponds here is very different from using a typical pond-resolving model. Such models solve pond and ice evolution on a grid, often in 3 dimensions, parameterizing the sub-grid scale processes that cannot be explicitly resolved. For example, in a recent attempt, \citet{skyllingstad2015simulation} solved a 3d model of pond evolution coupled with ice thermal evolution, assuming that ponds drain through the bulk of the ice, and allowing the possibility for an impermeable ice layer to form. Their model physics allowed for all stages of pond evolution to emerge naturally as the ice permeability evolves, but, to match the measured pond evolution, the ice permeability had to be tuned several orders of magnitude below the observed value. In this paper, for the first time, we explicitly model pond drainage through large holes and, as a result, we are able to match the observations using realistic physical parameters. Moreover, we use an explicitly solvable model that exhibits universality, which makes it numerically inexpensive and easily interpreted. 

This paper is organized as follows: First we formulate a model of drainage through large holes. Next, we use this model to explain why the ponds organize around the percolation threshold. After this, we show that the drainage process is in fact universal. In the two following sections, we use universality to formulate an equation for pond coverage evolution that approximately solves the hole model. Next, we show that our results are consistent with observations. Then, we discuss the dependence of pond coverage evolution on physical parameters and the challenges of pond modeling and, finally, we conclude with a summary of our results. All of the parameters used in the paper are summarized in table \ref{tab:1}. We summarize technical details required to reproduce the results of the paper in the Supporting Information (SI). 


\section{The hole model}\label{sec:hole_model}

To investigate pond drainage through macroscopic holes, we created a simple model of pond drainage, sketched in Fig. \ref{fig:1}b and explained in detail in SI section \ref{sec:model}. We discuss the physics of hole formation later, in section \ref{sec:stageIIevol}. \linelabel{B2.1}To mimic the conditions at the end of stage I of pond evolution, we start with a random synthetic two-dimensional ice surface, a large fraction of which is covered by water. \linelabel{B6.2}We are assuming that variations in ice topography are initially much smaller than freeboard height (mean height above sea level), such that no regions of the ice are initially below sea level. Such a configuration is consistent with conditions on undeformed first-year ice. The ice topography can change over the course of the melt season, as preferential melting of ponded ice alters the topography. A hole opens at a random location on the ice surface and drains water from all of the ponds it is connected to. As drainage through this hole progresses, some parts of the surface can become disconnected from the hole and can no longer drain through it. Drainage stops when either the hole is exposed to the atmosphere or when the pond connected to the hole reaches sea level. As water is lost from the ice surface, the ice floats up to maintain hydrostatic balance. Hydrostatic balance depends on ice thickness, which can decrease over time. Since ponded ice melts faster than bare ice, we evolve the topography by preferentially melting ponded regions. We are assuming that this melt occurs much slower than drainage through a hole. \linelabel{A11}\linelabel{B1} We are also assuming that the topography only changes due to albedo differences between bare and ponded ice so that channels that connect ponds cannot form as the ponds drain. Later, more holes open on the ice surface and the process continues until all the ponds are at sea level. 

\linelabel{B1.2}In practice, we implemented our model on a square grid, typically $500 \times 500$ grid points in size, with boundaries that we regard as rigid walls over which ponds cannot spill. We initiated the model by randomly generating a topography according to one of the schemes described in SI section \ref{sec:topos} and setting an initial water level, typically so that 100\% of the surface is water-covered. During each simulation run, we tracked the ice height above sea level and the water level at each grid point as well as a global variable, $\theta$, loosely interpreted as ice temperature. $\theta$ increases at a fixed rate and determines where the holes open - to each grid point, we assign a critical value, $\theta_0$, drawn independently from a normal distribution, so that a hole opens at that grid point when $\theta > \theta_0$. Ice thickness, which controls the height of the freeboard, decreases at a constant rate $\text{d}H/\text{d}t$, which we take to be an independent model parameter, separate from the surface melt rate. We consider both simulations with $\text{d}H/\text{d}t = 0$ and with $|\text{d}H/\text{d}t| > 0$. We simulate pond and ice evolution iteratively - 1) first, we update $\theta$ to find grid points where the new holes open, 2) then we drain the ponds through these holes by incrementally decreasing the water level, at each step updating the ponds connected to the holes until either the holes emerge above the pond surface or ponds fall to sea level, 3) next, we update the surface topography by preferentially melting ponded regions, and 4) finally, we update ice thickness and impose hydrostatic balance. We repeat these steps until $\theta$ is high enough so that all grid points have a hole in them. We describe this model in more detail and carefully consider all of the model assumptions in SI sections \ref{sec:model} and \ref{sec:assumptions}. Code for this model is archived in \citet{pedjapopovic20203930528} and is also available at \url{https://github.com/PedjaPopovic/hole-model}. 


Ice topography determines how the ponds will drain. As we will show, the effects of topography on pond coverage can be summarized with only a few parameters. Importantly, though, for our analysis to apply, ice topography has to be well-described by a single length scale, $l_0$, defined, for example, as the autocorrelation length. In the case of melt ponds, $l_0$ would be determined by the typical size of snow dunes. By only considering single-scaled topographies, we are restricting our analysis to flat regions of the ice away from large-scale features such as the ridges. \linelabel{A10} As we mentioned in the introduction, an especially important property of the surface is its percolation threshold, $p_c$. We estimate $p_c$ for a given topography as the coverage fraction at which a connected ``pond'' that spans the domain first appears as we incrementally raise the ``water level'' (a horizontal plane that cuts through the topography, see SI section \ref{sec:topos} for details). Motivated by the void model of \citet{popovic2018simple} where ponds surround circular ``snow dunes,'' in \citet{popovic2020snow} we developed a ``snow dune'' model of the ice topography generated by summing randomly placed mounds of Gaussian form, each with a randomly chosen horizontal scale and a height proportional to that scale. There, we showed that this topography reproduces the statistical properties of the pre-melt snow surface highly accurately. An example of this surface is shown in Fig. \ref{fig:1}c and a typical configuration of ponds after drainage through one hole on this surface is shown in Fig. \ref{fig:1}d. However, our analysis applies to any random surface with a single characteristic scale, and we also show the results for other topography types. We discuss all of these topographies in SI section \ref{sec:topos}.

\section{The origin of the percolation threshold in melt ponds}

\begin{figure*}
\centering
\includegraphics[width=1.\textwidth]{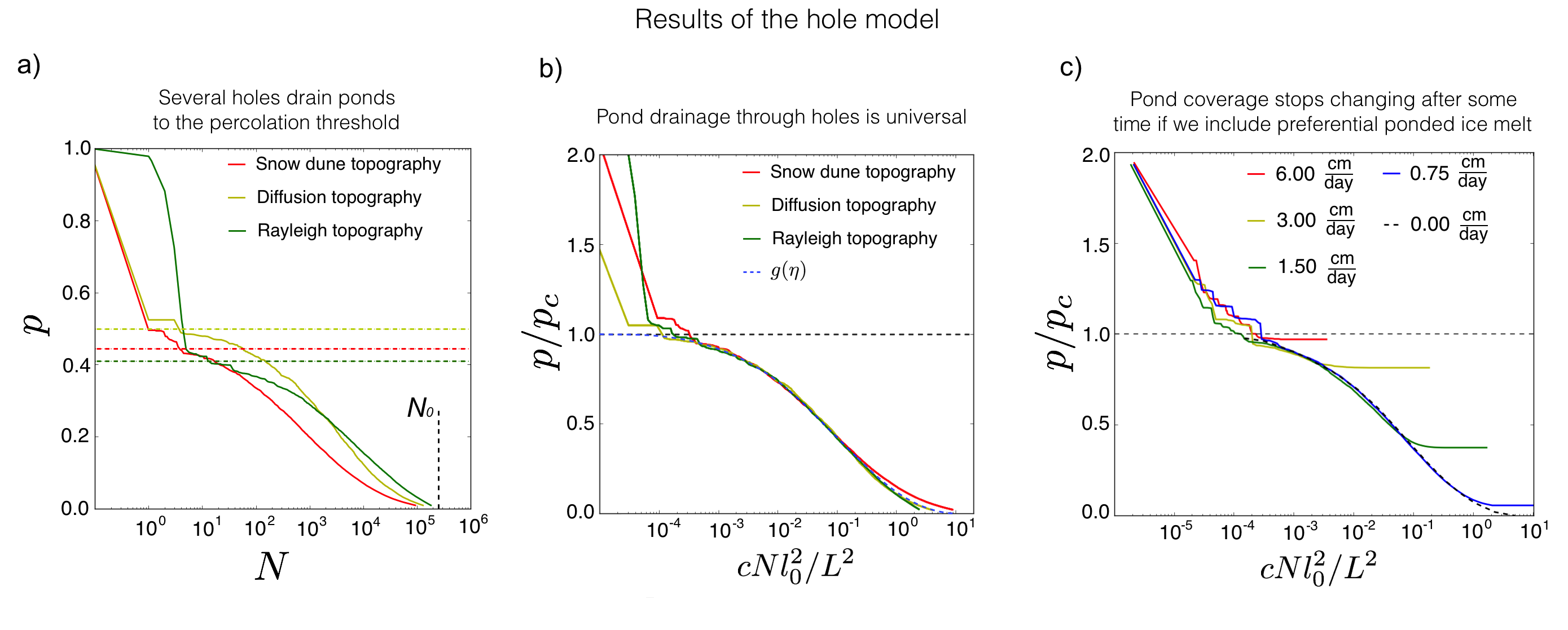}
\caption{ a) Semi-log plot of pond coverage fraction as a function of the number of open holes for three different topography types and no ice melt. Percolation thresholds for each of the three topographies are shown as horizontal dashed lines. The vertical dashed line marks the total number of pixels in these simulations. The simulations start at $N = 0$ and $p=1$ which cannot be shown due to the logarithmic scale, so, for visual clarity, we artificially place the origin at $N = 0.1$. b) A plot of rescaled pond coverage, $\Pi \equiv p / p_c$, as a function of rescaled number of holes, $\eta\equiv c N l_0^2 / L^2$ for the three surfaces in panel a. The horizontal dashed line marks the percolation threshold, $\Pi = 1$. Following initial drainage above the percolation threshold, all three curves fall approximately on a universal function, $g(\eta)$, marked with a dashed blue line. c) Rescaled pond coverage, $\Pi$, as a function of rescaled number of holes, $\eta$, now with preferential ponded ice melt, for different rates of ponded ice melt relative to bare ice melt, $\text{d}h_\text{diff}/ \text{d}t$. Ice thinning rate is kept at 0, i.e., $\text{d}H/\text{d}t = 0$. The horizontal dashed line has the same meaning as in panel b.}
\label{fig:2}
\end{figure*}

Figure \ref{fig:2}a shows the pond coverage fraction, $p$, as a function of the number of open holes, $N$, in the hole model if we assume ice does not melt. Figure \ref{fig:2}a provides insight into the mechanism for pond coverage being organized close to the percolation threshold, $p_c$. Specifically, we can see that the first several holes drain the entire ice surface from $p = 1$ to $p \approx p_c$, while it takes on the order of $10^5$ more holes to drain the rest of the surface. In these simulations, the total number of pixels was $N_0 \approx 2.5\times 10^5$, so to fully drain the surface, a hole needed to exist on nearly every pixel. This resembles the opening of microscopic pathways in real ice, and explains why the transition from stage II to stage III is also marked with a transition from drainage through large holes to drainage through microscopic pathways. \linelabel{B13}This result was robust - we ran multiple simulations on three kinds of topographies and in each case the first several holes were able to drain ponds to below $p_c$, while the remainder of the drainage curves were robust against the randomness in our model.

We can now understand why the initial drainage is so abrupt and leads to pond coverage close to the percolation threshold. If pond coverage is above the threshold, then a large fraction of the surface is connected and a single hole can drain a vast area. On the other hand, if pond coverage is below the threshold, then ponds become disconnected and holes can drain ponds only incrementally. To explain why ponds remain close to the percolation threshold as more holes open, we need to include the fact that ponded ice melts faster than bare ice. In this case, the topographic variations that determine the pond patterns amplify over time, so that percolation pond patterns can be preserved if ponded ice melts sufficiently fast compared to bare ice. 

The percolation threshold is a statistical property of the ice surface. As such, it does not depend on many of the details of the topography. Importantly, it does not typically depend on dimensional properties such as the mean height or roughness (standard deviation), but rather on generic statistical properties of the surface \citep{weinrib1982percolation}. For example, any surface that has a height distribution with a point of symmetry (e.g., a Gaussian), will have $p_c = 0.5$ \citep{zallen1971percolation}.\linelabel{A12} Non-symmetric surfaces have a $p_c$ that deviates from 0.5, but our experiments on non-symmetric ``snow dune'' and Rayleigh topographies suggest that these deviations are typically not very large (see SI section \ref{sec:topos}). This means that the percolation threshold for real ice can likely be accurately estimated using an appropriate statistical model of ice topography.\linelabel{A11.2} We note that if channels that connect ponds form in a significant number as the ponds drain, statistical properties of the topography may change over the course of pond drainage, thereby lowering the value of the percolation threshold.

\section{Universality of drainage through holes}

\noindent
The curves for drainage on different topographies shown in Fig. \ref{fig:2}a all appear different. However, by rescaling $p$ and $N$, we can collapse these curves onto a single universal curve (Fig. \ref{fig:2}b). The appropriate rescaling is 
\begin{linenomath*}
\begin{eqnarray}
p & \rightarrow & \Pi \equiv \frac{p}{p_c} \label{eq:1} \\
N & \rightarrow & \eta \equiv c N \frac{l_0^2}{L^2} \text{ ,  } \label{eq:2} 
\end{eqnarray}
\end{linenomath*}
where $L$ is the size of the domain and $c$ is a non-dimensional number of order unity that, like $p_c$, depends on the type of the topography but not on its dimensional properties such as mean height or roughness. \linelabel{B53}Below we show that $c$ can be defined using a theoretical curve that arises from the general percolation theory. A factor $L^2/l_0^2$ in the parameter $\eta$ is approximately the number of ponds of size $l_0$ within the domain of size $L$. Therefore, the parameter $\eta$ represents, approximately, the number of open holes per pond of size $l_0$. As we can see in Fig. \ref{fig:2}b, the surface is significantly drained when $\eta$ is of the order one, meaning that it takes about one hole per pond to drain the surface. After this rescaling, drainage on all surfaces follows a universal law
\begin{linenomath*}
\begin{equation}
\Pi = g(\eta) \text{  . } \label{eqn:3}
\end{equation}
\end{linenomath*}
Pond coverage in our model falls on this universal curve after the first several holes have drained the ponds to the percolation threshold. For a general discussion on universality see \citet{goldenfeld1992lectures}.


\linelabel{B3}The universality of the curve $g(\eta)$ is a consequence of the universality of the percolation model near $p_c$. In SI section \ref{sec:g}, we use this fact to motivate the form of $g(\eta)$. In particular, we define a correlation function for ponds, $G(l)$, as the probability that two randomly chosen points on the surface, separated by a distance $l$, are both located on the same pond. With this definition, the integral of $G(l)$ is proportional to the fraction of the surface connected to a randomly located hole, and, so, can be related to an average decrease in pond coverage per hole. The universality of pond drainage then arises from the fact that, close to the percolation threshold, $G(l)$ has the same form for all models within the percolation universality class. In SI section \ref{sec:g}, we use this line of reasoning to show that $g$ is approximately a solution to an ordinary differential equation
\begin{linenomath*}
\begin{equation} \label{eqn:4}
\frac{\text{d} g}{\text{d} \eta} = - g^2 (1-g)^{-19/18} \text{  . }
\end{equation} 
\end{linenomath*}
This reasoning only holds near $p_c$, so, when $\Pi$ deviates significantly from 1, the universality breaks down, which we can see in the hole model with the slight deviation among the curves at low pond coverage in Fig. \ref{fig:2}b. We use Eq. \ref{eqn:4} to specify the constant $c$ in the parameter $\eta$. We chose $c$ so that the simulated curves $\Pi = g(\eta)$ best correspond to Eq. \ref{eqn:4}. All tested topographies yielded a similar scaling factor, $c$, so this constant had only a small effect in our simulations (see SI section \ref{sec:topos}). 

\linelabel{A13}Since $g(\eta)$ does not depend on model details, a solution for one model will apply to all models within the same universality class. Therefore, it is likely that solving for $g$ using Eq. \ref{eqn:4} will apply well to real sea ice, so long as real ice topography does not qualitatively differ from our synthetic surfaces. This is supported by \citet{popovic2020snow}, where we showed that pre-melt ice topography on undeformed ice is very accurately described by the synthetic ``snow dune'' surface we considered here. We note, however, that if ice topography changes significantly during drainage, e.g. due to the formation of channels that connect ponds, real ice may fall out of the percolation universality class, so more study is needed to ensure that our results apply to real ice. As a final remark, we note that the function $g(\eta)$ may in fact describe diverse physical phenomena, some of which may be seemingly very different from pond drainage on sea ice.

Identifying the function $g(\eta)$ is a central result of our paper. In the rest of the paper we will be concerned with using this function to understand pond evolution and its connection to measurable parameters. 

\section{Pond evolution during stage II} \label{sec:stageIIevol}

So far, we have neglected the preferential melt of ponded ice. In this section, we will include this effect and we will derive an equation for pond coverage time-evolution, $p(t)$, during pond drainage. In doing so, we will clarify the relationship between pond evolution and measurable physical parameters. 

Figure \ref{fig:2}c shows rescaled pond coverage as a function of the rescaled number of holes for different rates of ponded ice melt relative to bare ice melt, $\frac{\text{d}h_\text{diff}}{\text{d}t}  \equiv |\frac{\text{d}h_\text{pi}}{\text{d}t}| - |\frac{\text{d}h_\text{bi}}{\text{d}t}|$, where $ \frac{\text{d}h_\text{pi}}{\text{d}t}$ and $ \frac{\text{d}h_\text{bi}}{\text{d}t}$ are ponded and bare ice melt rates and $|...|$ is the absolute value. We are still neglecting the thinning of the ice, assuming that $\frac{\text{d}H}{\text{d}t} = 0$. We can see that, in this case, pond coverage still follows the same curve $p = p_c g(\eta)$ up to a point, $p = p_\text{min}$, when it stops changing as more holes open. As we hinted at before, the greater $\frac{\text{d}h_\text{diff}}{\text{d}t}$ is, the higher this pond coverage will be, and, if $\frac{\text{d}h_\text{diff}}{\text{d}t}$ is large enough, the pond coverage will get pinned to the percolation threshold. 

The pond coverage eventually stops changing because, after a certain amount of time, the base of the ponds melts below sea level so that new holes that open cannot drain the ponds fully. In this way, the pond patterns become ``memorized.'' This happens when ice melts through the thickness of the post-drainage freeboard, $h$. We can express $h$ and $\frac{\text{d} h_\text{diff}}{\text{d}t}$ in terms of physical parameters to find the time, $T_\text{m}$, to memorize the pond patterns
\begin{linenomath*}
\begin{eqnarray}
&h =  \frac{\rho_w - \rho_i}{\rho_w}  \frac{H}{1-p_\text{min}} \text{  ,}  \label{eqn:5} \\ 
&\frac{\text{d}h_\text{diff}}{ \text{d}t }= \frac{\Delta \alpha F_\text{sol}}{l_m} \text{  ,} \label{eqn:6} \\
&T_\text{m} \approx \frac{h}{\text{d}h_\text{diff} / \text{d}t} =  \frac{l_m}{\Delta \alpha F_\text{sol}} \frac{\rho_w - \rho_i}{\rho_w} \frac{H}{1-p_\text{min}} \text{  ,}  \label{eqn:7} 
\end{eqnarray}
\end{linenomath*}
where $H$ is the post-drainage ice thickness, $p_\text{min}$ is the post-drainage pond coverage, \linelabel{A14}$F_\text{sol}$ is the solar radiation flux, $\Delta \alpha$ is the albedo difference between ponded and bare ice, $l_m$ is the latent heat of melting in $\text{J}\text{m}^{-3}$, and $\rho_w$ and $\rho_i$ are the densities of water and ice. \linelabel{B14}Equation \ref{eqn:5} comes from hydrostatic balance of the ice floe, taking into account the fact that after drainage, only the non-ponded ice that covers a fraction $1-p_\text{min}$ of the total area is above sea level and can balance the buoyancy of the submerged ice. Equation \ref{eqn:6} follows from the assumption that bare and ponded ice melt differently only due to their albedo difference.

Based on the above, if we know the number of holes that have opened by time $t$, we can estimate the pond coverage evolution, as $p(t) \approx p_c g(\eta(t))$ for $t < T_\text{m}$ and $p(t) = p_\text{min} \equiv p_c g(\eta(T_\text{m}))$ for $t > T_\text{m}$. To do this, we must model the hole opening dynamics in some way. The formation of holes depends on ice microphysics, and is not very well understood. For this reason, we will first describe the hole opening process in a general way, making few assumptions. Afterwards, we will test concrete assumptions within this framework to make estimates of the pond coverage evolution.

\begin{figure*}
\centering
\includegraphics[width=1.\linewidth]{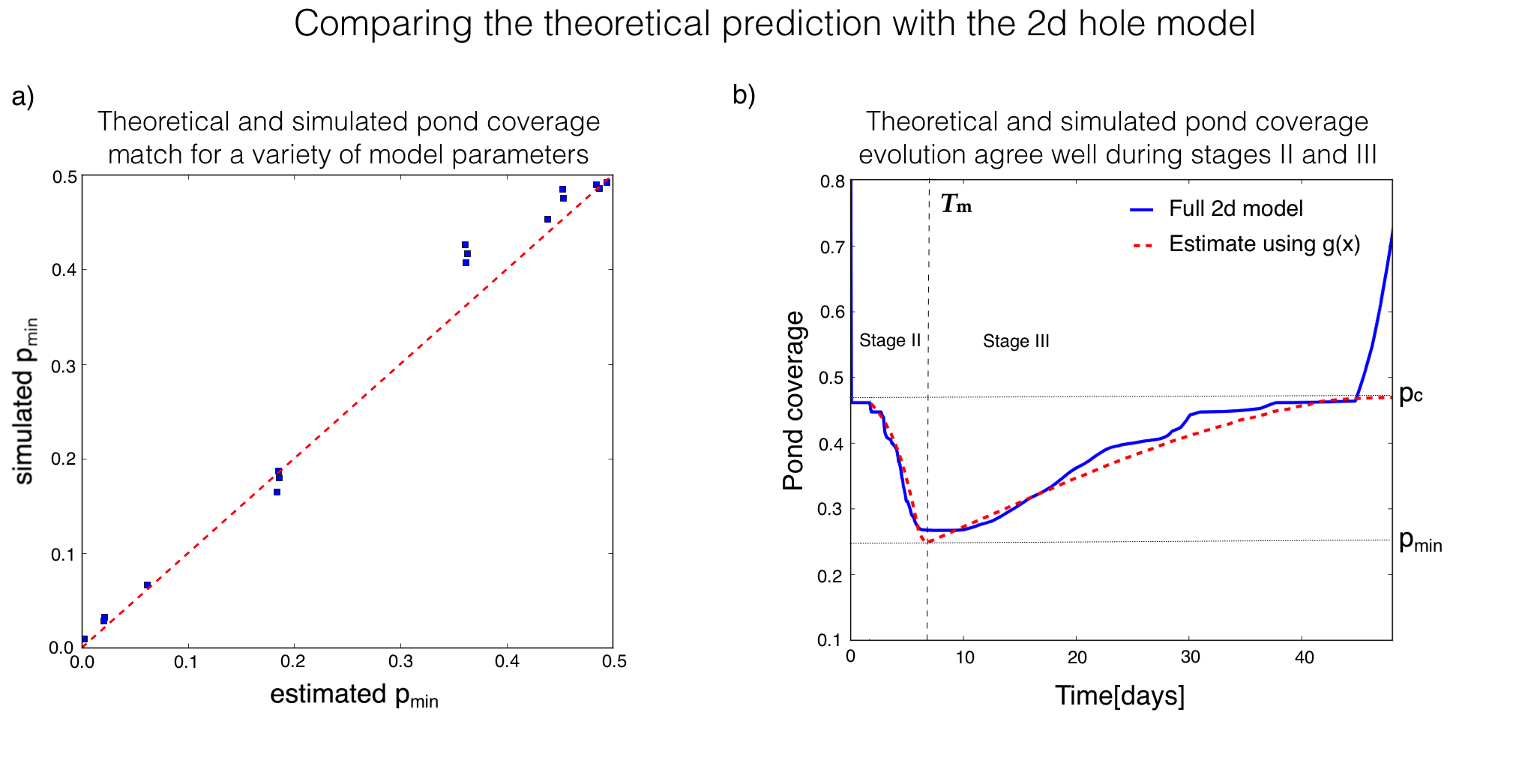}
\caption{a) Post-drainage pond coverage, $p_\text{min}$, found using our full 2d model against the coverage estimated using Eq. \ref{eqn:13}. Each point is a separate simulation with different rates of ponded ice melt and hole opening timescales. The red dashed line marks the 1:1 ratio. b) Pond coverage evolution using the full 2d model (blue line) and estimated using Eqs. \ref{eqn:12} and \ref{eqn:14} (red dashed line). Time of memorization, $T_\text{m}$, percolation threshold, $p_c$, and the post-drainage pond coverage, $p_\text{min}$, are marked with dashed and dotted black lines. Times earlier than $T_\text{m}$ correspond to stage II of pond evolution, while later times correspond to stage III. }
\label{fig:3}
\end{figure*}

Holes form mainly as enlarged brine channels \citep{polashenski2012mechanisms}. Such brine channels are tubes, roughly a centimeter in diameter, that run through the entire thickness of the ice \citep{cole1998observations}. \linelabel{B15}Not all brine channels span the entire depth of the ice \citep{lake1970salt}, so, likely, only some can become enlarged into holes. \citet{polashenski2012mechanisms} showed that, depending on the channel radius, ice temperature, salinity, and other bulk properties, a channel can either close or enlarge when meltwater flows through it. In the beginning all the channels are closed, but as the ice warms, some of them start to open. \citet{polashenski2017percolation} suggested that as the ice temperature passes through a particular threshold, some channels begin to open, while above a certain temperature nearly all channels become open. 

\noindent
Based on the above, we choose to model the evolution of the number of holes as
\begin{linenomath*}
\begin{eqnarray}
&N(t) =N_0 F\big( \frac{t-t_0}{T_\text{h}} \big) \text{ ,  }\label{eqn:8} \\
& t_0 \sim - T_\text{h} F^{-1} ( \frac{1}{N_0} )  \text{ ,  }\label{eqn:9} 
\end{eqnarray}
\end{linenomath*}
where $N_0$ is the total number of brine channels that can possibly become holes, $F$ is the fraction of those channels that become holes by time $t$, and can be seen as a cumulative distribution of some underlying probability density function $f$, and $T_\text{h}$ and $t_0$ are the width and the center of this distribution. \linelabel{A37}The parameter $t_0$ can be interpreted as the time between when the first hole opens and when a fraction of holes, $F(0)$, open. Equation \ref{eqn:9} then follows by noting that time $t=0$ corresponds to the opening of the first hole, so that $N(t = 0) = 1$. This relationship is approximate because the timing of opening the first hole is intrinsically stochastic. So, each independent model run will have a slightly different $t_0$ even if all large-scale parameters are the same (see SI section \ref{sec:holedist}). Nevertheless Eq. \ref{eqn:9} shows that, to first order, $t_0$ is not an independent parameter. Currently, we are only assuming that the distribution $F$ has a well-defined width, controlled by a unique hole opening timescale, $T_\text{h}$. Thus, we are assuming that a significant fraction of holes open within time $T_\text{h}$, and that within several $T_\text{h}$ almost all of the brine channels open. From Eq. \ref{eqn:8}, we can see that under these relatively broad assumptions, ice microphysics contributes to pond evolution by changing the hole timescale, $T_\text{h}$, number of channels, $N_0$, and the distribution $F$. 

The hole opening timescale, $T_\text{h}$, depends on specific mechanisms that control the formation of holes and is, therefore, more difficult to relate to measurable parameters than the memorization timescale (Eq. \ref{eqn:7}). \citet{polashenski2017percolation} suggested that a significant fraction of holes open up when the ice interior warms by some amount $\Delta \theta \sim 1 ^\circ\text{C}$ beyond the temperature at which the first hole opens. This suggests a way to relate $T_\text{h}$ to physical parameters in a way that is consistent with observations, but we note that a better understanding of hole formation physics is needed to make this estimate more realistic. Therefore, we estimate 
\begin{linenomath*}
\begin{eqnarray}
& T_\text{h} \approx \frac{\Delta \theta}{\text{d} \theta / \text{d} t} \text{ , }\label{eqn:10} \\
& \frac{\text{d} \theta}{\text{d} t} \approx \frac{\theta_0^2 }{\rho_i \gamma S} \Big( c^* k \frac{\theta_0}{H^2}  + (1-\alpha_p) F_\text{sol} \kappa e^{-\kappa z^*} \Big) \text{  ,}\label{eqn:11} 
\end{eqnarray}
\end{linenomath*}
where $\text{d} \theta / \text{d} t$ is the warming rate of the ice interior, $S$ is the ice salinity, $\theta_0$ is a reference temperature in degrees Celsius at which the holes tend to start opening, $\gamma$ is a constant that relates the ice heat capacity to the salinity, $k$ is the thermal conductivity of the ice, $\alpha_p$ is the albedo of ponded ice, $\kappa$ is the extinction coefficient from Beer's law, $z^*$ is the depth within the ice at which we are estimating the warming rate, and $c^*$ is a constant that accounts for the shape of the vertical temperature profile. Equation \ref{eqn:11} is an order-of-magnitude estimate of the ice warming rate in terms of measurable parameters that we derive in SI section \ref{sec:heat} following \citet{bitz1999energy}. We note, however, that $T_\text{h}$ can change by a factor of several by changing the under-constrained properties such as the depth at which the freshwater plugs form or the reference temperature, $\theta_0$. 

\noindent
With a model for hole opening, we can estimate the pond evolution, $p(t)$, and pond coverage after drainage, $p_\text{min}$, by combining Eqs. \ref{eqn:3} and \ref{eqn:8}
\begin{linenomath*}
\begin{eqnarray}
&  p(t) \approx p_c g\left(c N_0  \frac{l_0^2}{L^2} F \left( \frac{t - t_0}{T_\text{h}}\right)  \right)   \text{ , }  &t < T_\text{m} \label{eqn:12} \\
& p(t) \approx p_\text{min} \equiv p(T_\text{m})  \text{ , }  & t > T_\text{m} \text{   .  } \label{eqn:13} 
\end{eqnarray}
\end{linenomath*}
These equations can be solved once we assume a concrete distribution $F$. As a default, we used a cumulative normal distribution $F$ throughout the paper. In SI section \ref{sec:holedist}, we explored other distributions and showed that, due to the fact that $N_0$ is very large, pond evolution only depends on the tail of $F$. Since $T_\text{m}$ depends on $p_\text{min}$ through Eq. \ref{eqn:7}, Eqs. \ref{eqn:7} and \ref{eqn:13} have to be solved simultaneously for $p_\text{min}$ and $T_\text{m}$. To test the above relations, we compared $p_\text{min}$ found using Eq. \ref{eqn:13} against our full 2d hole model (see SI section \ref{sec:compare} for details). Figure \ref{fig:3}a shows that the simulations and the estimates agree well, confirming that the above equations do in fact approximately solve the full hole model. In SI section \ref{sec:params}, we use Eqs. \ref{eqn:7} and \ref{eqn:9} - \ref{eqn:13} to explore how the pond coverage depends on physical parameters.

\section{Pond evolution during stage III}\label{sec:stage3}

\begin{figure*}
\centering
\includegraphics[width=0.65\linewidth]{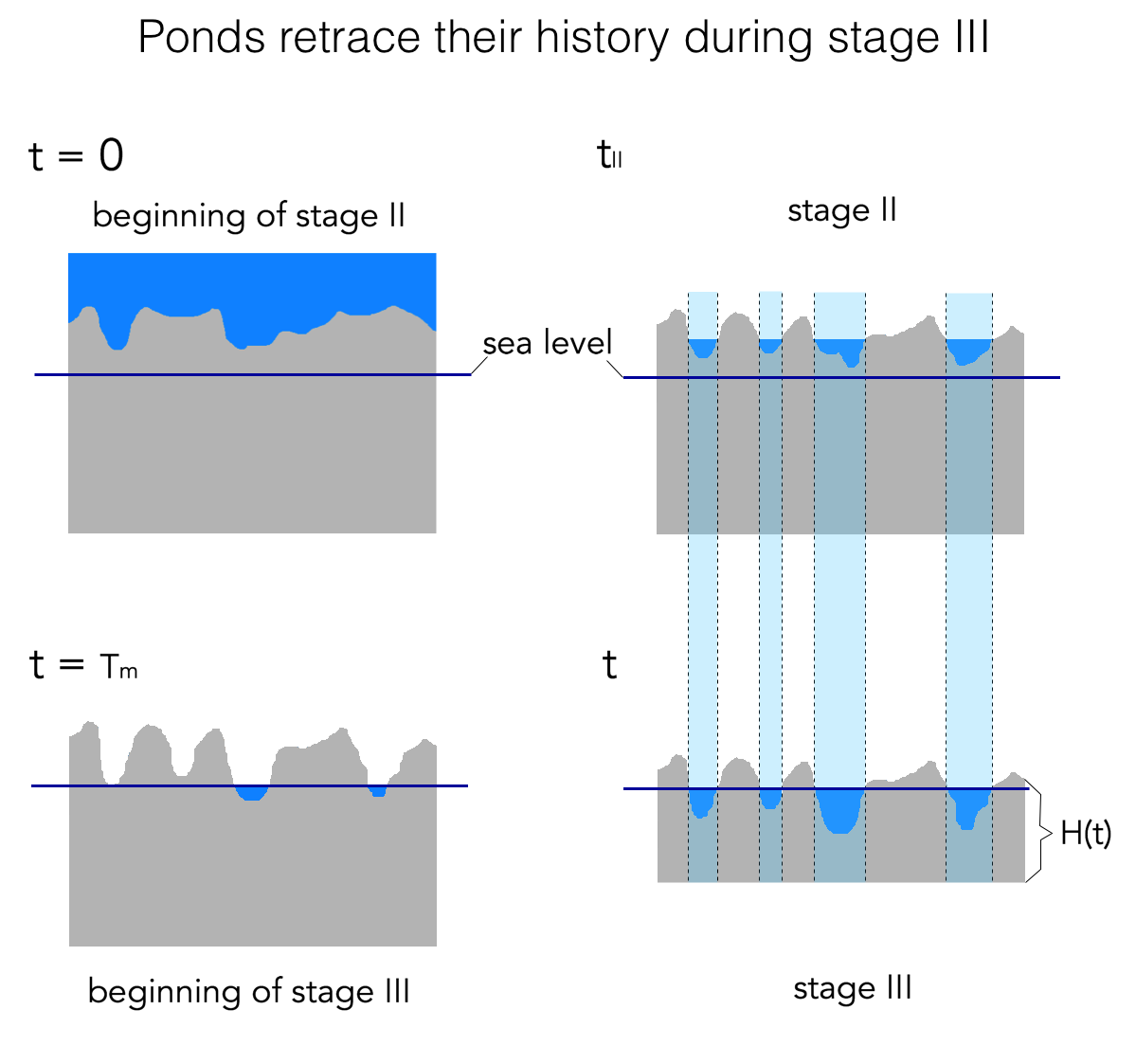}
\caption{A schematic of pond evolution under the assumptions stated in section \ref{sec:stage3}. Stage II begins at time $t = 0$, with a high pond coverage and ice surface above sea level. At a time $t_\text{II}$ during stage II, water level decreases and pond coverage evolves according to Eq. \ref{eqn:12}. We assume that the ponds are approximately level so their boundaries define a level-set of the topography. Ponded ice may melt in a complicated way, but we assume that bare ice melts spatially uniformly, so the topography left in the wake of the decreasing water level remains unchanged. When ponds reach sea level, stage II ends and stage III begins. During stage III, ponds are defined by the intersection of sea level and the ice topography, and are thus also level-sets of the topography. The above-sea-level topography at the beginning of stage III determines pond evolution during stage III if there is no lateral melt of pond walls. As the ice thins, sea level intersects this topography at different levels. Since this topography has not changed since the last time it was ponded during stage II, there exists a time $t$ during stage III for which the horizontal distribution of ponds is the same as it was at a time $t_\text{II}$ during stage II. In this sense, ponds retrace their history. For visual clarity, we have not shown ice thinning during stage II, but it is assumed that ice thins throughout its evolution.}
\label{fig:5}
\end{figure*}

We have shown that the universal function $g(\eta)$ can be used to solve our hole model, providing a formula for pond coverage evolution during stage II. In our model so far, we have neglected ice thinning, and so were unable to explicitly model stage III of pond evolution. Here we will extend our analysis to this stage as well. We will show that the same function $g(\eta)$ also governs the evolution of pond coverage during stage III under certain assumptions. 

\linelabel{A15}\linelabel{B7}We start by explaining the relationship between pond behavior during stage II and stage III (see Fig. \ref{fig:5}). To this end, we have to make several assumptions. In particular, we will assume that ponds are approximately level during stage II,  that bare ice melts at a spatially uniform rate, that pond coverage only decreases during stage II and only increases during stage III, and that there is no lateral melt during stage III. If ponds during stage II are approximately level, ice at the pond boundaries has approximately the same height, and is thus, approximately, a level-set of the ice topography. As the ponds drain and the water level decreases during stage II, ice at the pond boundaries quickly becomes pond-free and, assuming that bare ice melts at a spatially uniform rate, remains approximately level for the remainder of stage II. During stage III, ponds are drained to sea level nearly completely, so pond boundaries are again level-sets of the ice topography and the pond coverage is equal to the fraction of the ice surface below sea level. If there is no lateral melting of pond walls, pond growth during stage III is entirely due to ice thinning and consequent freeboard sinking, which leads to more of the ice surface falling below sea level. Thus, the above-sea-level topography, which consists only of bare ice, does not change until it is submerged. This means that pond coverage at any time during stage III is uniquely determined by the above-sea-level topography at the beginning of stage III, which is unchanged since it was last ponded during stage II, and the current ice thickness, $H(t)$, which determines the freeboard height. Therein lies the connection between stage II and stage III ponds - both are approximately level-sets of the same topography created in the wake of decreasing water level during stage II. In this sense ponds during stage III approximately retrace their history - \linelabel{B8}for every water level during stage II there exists a corresponding ice thickness during stage III so that ponds are approximately the same. Using this idea, we can estimate the pond evolution during stage III. 

In order for ponds to retrace their history, the assumptions we stated at the beginning of the previous paragraph have to hold. These assumptions are likely satisfied in reality, at least approximately. In our hole model, ponds are certainly not exactly level during stage II, since this would prevent them from becoming disconnected. Nevertheless, this assumption likely holds well enough so that ponds approximately retrace their history, as we show below by directly simulating stages II and III in our model. Note that we only had to assume that bare ice melts spatially uniformly, and made no assumptions about ponded ice - whichever way the ponded ice melts, the topography set during stage II will be re-submerged during stage III, and ponds will retrace their history. However, a spatially non-uniform melt rate of ponded ice may affect the mapping between stage II and stage III ponds, so, to construct this mapping explicitly, below we will consider only the case where ponded ice does melt uniformly in both time and space. 

Now, we can proceed to explicitly relate stage II and stage III pond coverage. Consider a time $t_\text{II}$ since the beginning of stage II at which the pond coverage is $p_\text{II}(t_\text{II})$. \linelabel{A15.2}\linelabel{B9}If the ice is flat compared to freeboard thickness at the beginning of stage II and ponded ice melts uniformly, the depth of topographic depressions created by preferential ponded ice melting by time $t_\text{II}$ is approximately $\delta h(t_\text{II}) \approx \frac{\text{d}h_\text{diff}}{\text{d}t} t_\text{II}$. Thus, ice that was ponded at $t_\text{II}$ will have carved depressions of at least $\delta h$ by the beginning of stage III. On the other hand, since pond coverage only decreases during stage II, any non-ponded location at $t_\text{II}$ will have carved depressions less than $\delta h$ deep by the beginning of stage III. During stage III, depressions of depth $\delta h$ will be below sea level when the freeboard height, $h$, is less than $\delta h$. So, if the ponds retrace their history, we can use the relationship $h(t) = \delta h(t_\text{II})$, to find the time $t_\text{II}$ at which the pond coverage was the same as it is at time $t$ during stage III. Thus, we find $t_\text{II}(t) \approx \frac{h(t)}{\text{d}h_\text{diff}/\text{d}t}$.

Note that $t_\text{II}$ has the same form as the memorization timescale, $T_\text{m}$, given by Eq. \ref{eqn:7} - both are a ratio of a freeboard thickness, $h$, and a differential melt rate, $\text{d}h_\text{diff}/\text{d}t$, the only difference being that Eq. \ref{eqn:7} is estimated using $h$ at the end of stage II and $t_\text{II}$ is estimated using $h$ at some time, $t$, during stage III. So, we can use $T_m(t)$ calculated by Eq. \ref{eqn:7} using the values for ice thickness and pond coverage at time $t$, and define $t_\text{II}(t) = T_\text{m}(t)$. We thus relate stage II and III pond coverage as $p(t) = p_\text{II}(T_\text{m}(t))$. Therefore, using Eq. \ref{eqn:12}, the pond evolution during stage III can be approximately captured as
\begin{linenomath*}
\begin{eqnarray}  
& p(t)  \approx p_c g\Big(c N_0 \frac{l_0^2}{L^2}  F\Big(\frac{T_\text{m}( t ) - t_0}{T_h}\Big) \Big)  \text{  , } \label{eqn:14} \\
& T_\text{m}(t) \approx \frac{h(t)}{\text{d}h_\text{diff} / \text{d}t} =  \frac{l_m}{\Delta \alpha F_\text{sol}} \frac{\rho_w - \rho_i}{\rho_w} \frac{H(t)}{1-p(t)} \text{  .}  \label{eqn:15} 
\end{eqnarray}
\end{linenomath*}
This equation is applicable after complete pond drainage to sea level. In SI section \ref{sec:compare}, we note that the transition between stage II to stage III can be estimated as the time at which $p(t)$ approximated using Eq. \ref{eqn:14} exceeds $p(t)$ estimated using Eq. \ref{eqn:12} (i.e., when $t > T_\text{m}( t )$). After choosing $F$, Eqs. \ref{eqn:12} and \ref{eqn:14} together describe the pond evolution after the beginning of pond drainage below the percolation threshold. 

We test these equations against our full 2d hole model that includes ice thinning and assumes a constant thinning rate, $\text{d}H/\text{d}t$, in Fig. \ref{fig:3}b (see SI section \ref{sec:compare} for more details). We can see that the two agree well most of the time, suggesting that our assumption about ponds retracing their history is justified in our model. Since $g \leq 1$, Eqs. \ref{eqn:12} and \ref{eqn:14} predict that $p_c$ is an upper bound on pond coverage during stages II and III. This is approximately obeyed in the full hole model where the pond coverage during stage II quickly falls below the percolation threshold after the first several holes, and where the ice quickly floods during stage III after pond coverage reaches $p_c$. The rapid flooding after the pond coverage reaches $p_c$ during stage III in the model is due to the fact that we assumed the ice was very flat at the beginning of stage II. In reality, this assumption may not strictly hold, but we still expect the percolation threshold to be an approximate upper bound after which flooding, and subsequent ice disintegration, follows more rapidly than before.

\section{Comparison with observations}

\begin{figure*}
\centering
\includegraphics[width=1.\textwidth]{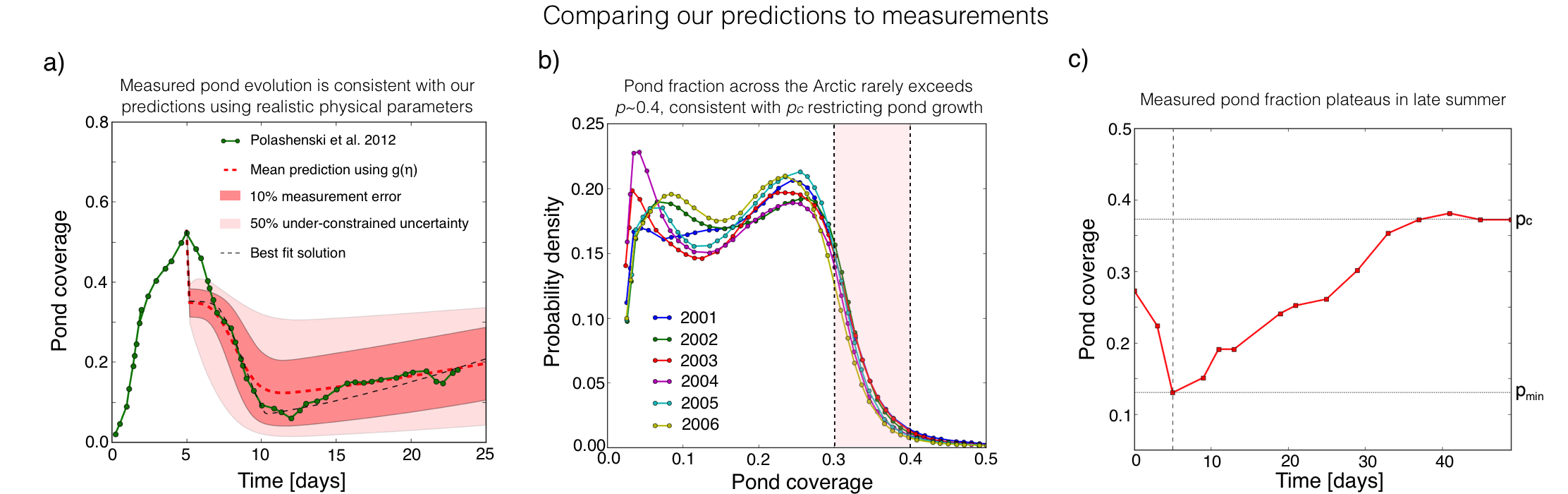}
\caption{a) The green line represents time evolution of pond coverage measured by Polashenski et al. (2012) near Barrow, Alaska in the summer of 2009. The red dashed line and the dark shaded region are the mean and one-standard-deviation uncertainty of an ensemble of pond coverage evolutions found using Eqs. \ref{eqn:12} and \ref{eqn:14}. For each pond evolution within the ensemble, every physical parameter that enters the pond evolution equations was selected randomly from a normal distribution with a mean estimated using measurements by Polashenski et al. (2012) and a standard deviation equal to 10\% of the mean. The black dashed line is the least-squares best fit prediction using parameters that are within the 10\% measurement error. The light shaded region is the one-standard-deviation uncertainty region assuming a 10\% measurement error for all the parameters directly measured by Polashenski et al. (2012) and a 50\% uncertainty for under-constrained microphysical parameters such as the shape parameter for the temperature profile, $c^*$, the reference temperature, $\theta_0$, and the depth at which the freshwater plugs form, $z^*$. Values of 10\% and 50\% are not estimates of real uncertainties, but were chosen simply to demonstrate the sensitivity to model parameters. We note that the slight increase in the upper boundary of the light shaded region during stage II simply corresponds to increased uncertainty during that time, and no actual pond coverage trajectory increases during stage II. b) Distribution of pond coverage across the Arctic for different years derived from MODIS satellite data \citep{MODIS_v02}. The frequency of observations declines rapidly between $0.3 < p < 0.4$ and very few observations show $p>0.4$, consistent with our predictions. The light shaded region shows the likely values for the percolation threshold estimated by \citet{popovic2018simple}. c) Observations of pond coverage evolution made along a transect during the 1998 SHEBA mission. The data are taken from \citet{perovich2003thin}. Horizontal and vertical lines mark the estimated minimum pond coverage, $p_\text{min}$, percolation threshold, $p_c$, and the timing of the transition between stages II and III. }
\label{fig:4}
\end{figure*}

We can now proceed to compare our model predictions to observations. We describe the details of these comparisons in SI section \ref{sec:data}. \citet{polashenski2012mechanisms} collected extensive field data on pond coverage and ice properties during the summer of 2009 near Barrow Alaska. Using their measurements and Eqs. \ref{eqn:7} and \ref{eqn:9} - \ref{eqn:11}, we were able to estimate all of the parameters that enter Eqs. \ref{eqn:12} and \ref{eqn:14}. Figure \ref{fig:4}a shows the measured pond coverage, along with our predictions and an estimated margin of error due to uncertainty in the physical parameters that enter the pond evolution equations, while Table \ref{tab:Results} summarizes our estimates of the timescales $T_\text{m}$ and $T_\text{h}$, the minimum pond coverage, $p_\text{min}$, and the percolation threshold, $p_c$, as well as their margin of error. Since so many physical parameters contribute to pond evolution, even modest uncertainty in measurements leads to relatively large uncertainty in pond coverage. In fact, we find that to match the pond coverage evolution to within 10\% all of the physical parameters would have to be known to within about 1\%. \linelabel{A33}\linelabel{B70}Equation \ref{eqn:12} predicts that ponds drain instantly to the percolation threshold as soon as the first hole opens. The measurements, on the other hand, show a more gradual decline in pond coverage towards the percolation threshold, likely due to the fact that real holes take time to grow and cannot drain ponds instantly. \linelabel{B17}Nevertheless, we can see that observations are consistent with our predictions below the percolation threshold, and our equations capture the pattern of pond coverage variability over time. We can see that within the limited range of reasonable physical parameters, we are able to choose a combination that predicts a pond evolution that accurately matches observations below $p\approx 0.36$. 

\begin{table}
\centering
\setstretch{1.}
\caption{Estimates of the timescales $T_\text{m}$ and $T_\text{h}$, the minimum pond coverage, $p_\text{min}$, and the percolation threshold, $p_c$, found using Eqs. \ref{eqn:7}, \ref{eqn:10}, \ref{eqn:11}, and \ref{eqn:13}. The explored range represents the minimum and maximum estimate of these parameters over an ensemble of pond coverage evolutions corresponding to the dark shaded region in Fig. \ref{fig:4}a, while the best fit estimates correspond to the least-squares best-fit pond evolution curve over this ensemble shown as the black dashed line in Fig. \ref{fig:4}a. }
\begin{tabular}{| c | c | c | c | c | }
\hline
& $T_\text{m}$ [days] & $T_\text{h}$ [days] & $p_\text{min}$ & $p_c$ \\
\hline		
Best fit & 4.4 & 2.0 & 0.1 & 0.36 \\
Explored range & (2.3, 9.2) & (1.0,6.6) & (0.0,0.3) & (0.2,0.5) \\
\hline
\end{tabular}
\label{tab:Results}
\end{table}

\linelabel{B11.3}As we discussed before, our model predicts an approximate upper bound on pond coverage during stages II and III. In particular, we showed that after the beginning of stage II, ponds quickly drain to $p_c$, while at the end of stage III, the entire floe quickly floods when pond coverage exceeds $p_c$. This means that ponds likely spend little time at coverage higher than the percolation threshold. Moreover, this approximate upper bound may also be present during stage I. If during stage I there exist large flaws in the ice such as cracks, holes, or even the floe edge, that can quickly drain large volumes of water, the percolation threshold would also represent an upper limit on pond coverage during this stage. If such flaws exist, the pond coverage would in fact be bounded throughout the entire melt season. In Fig. \ref{fig:4}b, we show the pond coverage distribution across the entire Arctic estimated from MODIS (Moderate Resolution Imaging Spectroradiometer) satellite data \citep{MODIS_v02}. The frequency of pond coverage observations is relatively high for $p < 0.3$ and declines rapidly for higher pond coverage values. Moreover, the pond coverage very rarely exceeds 0.4, consistent with our prediction that an upper bound on pond coverage exists. Values between 0.3 and 0.4 are also consistent with \citet{popovic2018simple} who estimated $p_c$ directly from pond photographs, and found it to be roughly between 0.3 and 0.4. \linelabel{A19}\linelabel{B11.2}We note that these values for the percolation threshold are about 0.1 lower than $p_c$ predicted by the ``snow dune'' topography, developed in \citet{popovic2020snow} (and described in SI section \ref{sec:topos}), that accurately matches the pre-melt ice topography and predicts a $p_c$ between 0.4 and 0.5. This decrease could be due to processes such as the formation of connecting channels between ponds during pond drainage that change the topography and increase the efficiency of drainage. 

\linelabel{B11}In the previous section and in Fig. \ref{fig:3}b, we showed that our model predicts that the pond coverage will approach $p_c$ in the late stages of pond evolution, after which the ice rapidly floods, presumably leading to disintegration. This late-season approach to $p_c$ and subsequent rapid flooding may in fact account for the observations by \citet{popovic2018simple} that late summer ponds are organized close to the percolation threshold. \linelabel{A18}Furthermore, observations made during the 1998 SHEBA mission, reported in \citet{perovich2003thin}, and shown in Fig. \ref{fig:4}c, showed stage III pond evolution that strongly resembles our model prediction that pond growth slows down as it approaches $p_c$. There, starting from a minimum, the pond coverage increased over a period of roughly 40 days approaching a point ($p \approx 0.37$) after which it approximately stopped changing for the remainder of the field experiment (compare Figs. \ref{fig:3}b and \ref{fig:4}c). Again, this value of $p \approx 0.37$ is a plausible percolation threshold according to \citet{popovic2018simple} and our previous discussion in this section.

\section{Discussion}\label{sec:discussion}

We now discuss how pond coverage in our model depends on physical parameters, how our model might be incorporated into a large-scale sea ice scheme, and the challenges of pond modeling. In the previous sections, we have related the pond evolution to parameters $p_c$, $H$, $\alpha_i$, $\alpha_p$, $F_\text{sol}$, $S$, $\Delta \theta$, $\theta_0$, $z^*$, $N_0$, $L$, and $l_0$, all of which can be estimated by measurements, at least in principle. However, not all of these parameters are easily estimated in a large-scale model and the qualitative dependence on some of them may not be appropriately captured in our parameterization due to the fact that we do not yet understand hole formation physics well enough. 

The timescale $T_\text{m}$ depends only on parameters that are available in large-scale models ($H$, $\Delta \alpha$, and $F_\text{sol}$). On the other hand, the hole opening timescale, $T_\text{h}$, depends on hole formation physics and is, thus, much more difficult to estimate. In our parameterization, $T_\text{h}$ is both a function of parameters available in large-scale models ($H$, $S$, $\alpha_p$, and $F_\text{sol}$) and those that can only be loosely constrained by field observations ($\Delta \theta$, $\theta_0$, and $z^*$). In addition to this, $T_\text{h}$ in our parameterization depends on salinity, $S$, and reference temperature, $\theta_0$, in a qualitatively counter-intuitive way (see SI section \ref{sec:params}). Namely, in our parameterization of $T_\text{h}$, we directly related the fraction of open holes to the ice interior temperature following the suggestion of \citet{polashenski2017percolation}. This led to $T_\text{h}$ being higher at higher $S$ and $\theta_0$, since these conditions imply a lower ice warming rate. This is counter-intuitive because higher salinity and reference temperature also imply a higher brine volume fraction which, in reality, likely increases the hole formation rate, lowering $T_\text{h}$. For this reason, we believe that the assumption of using the ice temperature to diagnose hole opening may need to be reconsidered. \linelabel{B44}Another parameter such as, for example, ice porosity, may be better suited to diagnose when holes in the ice begin to form. 

The dependence of pond coverage on most of the physical parameters in our model is qualitatively robust. An exception to this is the ice thickness. We find that the effect of thickness, $H$, depends qualitatively on the details of hole formation physics. Increasing $H$ has two effects - 1) it raises the freeboard, increasing $T_\text{m}$, and 2) it slows down the warming of the ice interior, increasing $T_\text{h}$. Therefore, the way in which $H$ affects pond coverage qualitatively depends on whether the effect on $T_\text{m}$ or $T_\text{h}$ dominates. \linelabel{A4.2} In SI section \ref{sec:params}, we show how a minor change in the assumptions about the hole formation physics may completely change the effect of ice thickness. We note that \citet{skyllingstad2015simulation}, who ran pond-resolving simulations that include pond drainage, did not find a systematic relationship between pond fraction and the initial ice thickness, consistent with our discussion here. Establishing the correct relationship between ice thickness and pond coverage may be important for estimating the strength of the ice-albedo feedback under global warming. To that end, our model could be used to test hypotheses about hole formation physics by making predictions about the relationship between pond coverage and physical parameters that can be tested against field data that exploit the natural variability in environmental conditions across the Arctic.

We believe that the dependence of pond coverage on all other physical parameters predicted by our model is qualitatively correct, even if it may be somewhat quantitatively inaccurate because of poor understanding of hole formation physics. In Fig. \ref{fig:params} of SI section \ref{sec:params}, we specifically look at the minimum pond coverage, $p_\text{min}$. There, we show that $p_\text{min}$ increases with the domain size, $L$, percolation threshold, $p_c$, ice albedo, $\alpha_i$, temperature range for hole opening, $\Delta \theta$, and the depth at which ice plugs tend to form, $z^*$. We show that it decreases with typical pond size, $l_0$, and pond albedo, $\alpha_p$. Finally, we show that it only weakly depends on the solar flux $F_\text{sol}$ and brine channel density, $n_0$. We note that prior to this investigation, it was not recognized that geometric parameters $L$, $l_0$, and $p_c$ can have any effect on pond coverage. 

\linelabel{A34}In addition to the uncertainties we discussed above, our model predicts that accurately estimating pond coverage evolution is a fundamentally difficult problem, regardless of how well we understand the ice and pond physics. Namely, our model predicts that pond coverage is highly sensitive to physical parameters, so even small errors in measurements or natural variability can lead to large variations in predicted pond coverage. In particular, as we have noted in the previous section, to be able to predict pond coverage to within several percent in our model, all of the physical parameters need to be known to within about 1\%. \linelabel{A37.2}Moreover, even if all of the bulk parameters are perfectly known, we still cannot perfectly constrain the pond evolution. Since the timing of the opening of the first drainage hole is intrinsically stochastic (see Eq. \ref{eqn:9} and SI section \ref{sec:holedist}), pond coverage will fluctuate from one situation to another even if all of the bulk parameters are identical. In our simulations this stochasticity in the timing of the first hole contributed to about $5\%$ variability in minimum pond coverage, $p_\text{min}$. 

Ice albedo is a critical parameter in large-scale models that controls the thermal evolution of the sea ice cover. Since melt pond coverage is a primary control on albedo, Eqs. \ref{eqn:12} and \ref{eqn:14} can be straightforwardly used as a physically sound and computationally inexpensive parameterization of the albedo evolution during stages II and III of pond evolution. In addition, these equations can be supplemented with a similarly inexpensive equation for stage I developed in \citet{popovic2020snow}. However, our discussion above clearly highlights two issues relevant for employing our model in a large-scale scheme - 1) ice microphysics that governs hole formation needs to be better understood, and 2) pond coverage possesses a built-in sensitivity to environmental conditions and stochasticity that greatly amplify any uncertainty that may exist in measurements. We emphasize that these challenges are not an artifact of our hole model - any model of melt ponds will need to address hole formation physics to accurately capture the dependence on physical parameters, and any model will likely face a similar sensitivity to physical parameters which simply stems from the fact that there are many parameters that control pond evolution. Our work here reveals that the natural way to parameterize pond coverage is through the number of drainage holes per characteristic area of the surface, $\eta$. Thus, if any model were able to track $\eta$ directly rather than break it down into a multitude of environmental parameters, uncertainty in estimates of pond coverage would greatly reduce. If such a reduction is not feasible, the low computational cost of our model could be exploited to assign a distribution of pond coverage at each grid point of a large-scale model rather than to provide a single pond evolution trajectory. In addition to the issues above that are likely a generic feature of melt pond physics, some of the simplifying assumptions are specific to our model, such as, for example, neglecting the formation of connecting channels between ponds or assuming that drainage through holes is instantaneous. These assumptions may also need to be further examined to make sure our model can be used as a reliable albedo parameterization.

\section{Conclusions}

\noindent
Our paper revolves around the observation that the percolation threshold is of special significance for melt ponds. We showed that this stems from the fact that ponds typically drain through large holes, making drainage easy above the threshold and difficult below. In this way the percolation threshold represents an approximate upper bound on pond coverage throughout most of, or, in some cases, the entire summer. The fact that pond coverage often lingers around the percolation threshold leads to universality that greatly simplifies this otherwise complex problem, and allows us to write a simple formula that describes pond evolution throughout most of the melt season. It also makes it possible to connect pond evolution with measurable parameters. Observations are consistent with all of our predictions. The formula for pond coverage we provided requires very little computational power. Therefore, it holds promise as a physical, accurate, and computationally inexpensive parameterization of pond coverage in large scale models. Finally, our work connects melt ponds with the broader field of critical phenomena and our results regarding the universality of drainage may be applicable to other systems whose evolution is governed by a critical point of a phase transition. \nocite{SupplementaryPercolationHoles}

\appendix

\begin{table*}[!ht]
\centering
\setstretch{0.6}
\small
\caption{A table of parameters used in the paper, their default values, and plausible ranges where it was possible to estimate the ranges.}
\begin{tabular}{| c | p{4.cm} | c | p{1.5cm} | p{1.25cm} | p{4.cm} |}
\hline
 & Name & Unit & Default value & Plausible range & Source\\
\hline		
$t$ & Time & day & & (0,30) &  \\	
$p$ & Pond coverage & None &  & (0,1) &  \\
$p_c$ & Percolation threshold & None & 0.35 & (0.3,0.5) & \citep{isichenko1992percolation,popovic2018simple}, this work\\
$p_\text{min}$ & Minimum pond coverage & None &  & (0,$p_c$) & \\
$\Pi$ & Renormalized pond coverage & None & & (0,1) &  \\
$l_0$ & Typical pond length-scale & m & 5.5 & (5.2,5.8) &  \citep{popovic2020snow} \\
$L$ & Drainage basin size & km & 1.5 &  & \citep{polashenski2012mechanisms} \\
$n_0$ & Density of brine channels & $\text{m}^{-2}$ & 100 & (60,120) & \citep{golden2001brine} \\
$N_0$ & Number of brine channels & None & $2.25\times 10^8$ &  & Inferred from $n_0$ and $L$ \\
$N$ & Number of open holes & None &  & (1,$N_0$) &  \\
$\eta$ & Renormalized number of open holes & None &  &  & this work \\
$c$ & Numerical drainage constant  & None & 3 & (3,4.1) & this work \\
$\rho_i$ & Density of ice & $\text{kg m}^{-3}$ & 900 &  & \\
$\rho_w$ & Density of water & $\text{kg m}^{-3}$ & 1000 &  & \\ 
$l_m$ & Latent heat of melting & $\text{kJ kg}^{-1}$ & $334$ &  & \\ 
$\gamma$ & Constant relating heat capacity to salinity & $\text{kJ kg}^{-1} \text{ppt}^{-1\circ} \text{C}$ & 18 &  & \citep{ono1967specific,bitz1999energy}\\ 
$k$ & Thermal conductivity & $\text{W m}^{-1\circ} \text{C}^{-1}$ & 1.8 & (1.3,2.0) & \citep{untersteiner1964calculations,bitz1999energy}\\ 
$\kappa$ & Extinction coefficient & $\text{m}^{-1}$ & 1.5 &  & \citep{untersteiner1961mass,bitz1999energy}\\ 
$H$ & Ice thickness & m & 1.2 & (0.5,3) & \\
$h$ & Post-drainage freeboard thickness & m & 0.1 & (0.05,0.3) & \\
$F_\text{sol}$ & Time-averaged solar flux & $\text{W m}^{-2}$ & 254 & (140,350) & \citep{polashenski2012mechanisms} \\
$\Delta \alpha$ & Albedo difference between pond and ice & None & 0.4 & (0.3,0.5) & \citep{polashenski2012mechanisms,perovich1996optical} \\
$\alpha_p$ & Pond albedo & None & 0.25 & (0.2,0.3) & \citep{perovich1996optical} \\
$S$ & Ice salinity & ppt & 3 &   & \citep{polashenski2012mechanisms} \\		
$\theta_0$ & Reference ice interior temperature & $^\circ$C & -1.2 &  & \citep{polashenski2017percolation} \\
$\Delta \theta$ & Temperature range for hole opening & $^\circ$C & 0.7 &  & \citep{polashenski2017percolation} \\
$c^*$ & Temperature profile effect on heat diffusion & None & 2 & (1,10) & this work \\
$z^*$ & Depth at which ice plugs form & m & 0.6 &  &   \citep{polashenski2017percolation}\\
$t_0$ & Center of the hole opening distribution & day & 13.4 &  & this work \\
$T_\text{m}$ & Memorization timescale & day & 4.4 &  & this work \\
$T_\text{h}$ & Hole opening timescale & day & 2.0 &  & this work \\
\hline
\end{tabular}
\label{tab:1}
\end{table*}

\newpage

\acknowledgments
We thank BB Cael and Stefano Allesina for reading the paper and giving comments. We thank Don Perovich for providing the melt pond image in Fig. \ref{fig:1}a. Predrag Popovi\'{c} was supported by a NASA Earth and Space Science Fellowship. This work was partially supported by the National Science Foundation under NSF award number 1623064. Code for the hole model is archived in  \citet{pedjapopovic20203930528} and is also available at  \url{https://github.com/PedjaPopovic/hole-model}. The authors declare no conflict of interest.

\bibliography{mybibliogJGR.bib}

\makeatletter\@input{xx_supp_percolation.tex}\makeatother
\end{document}


\renewcommand{\thetable}{S\arabic{table}}
\renewcommand{\thefigure}{S\arabic{figure}}
\renewcommand{\thesection}{S\arabic{section}}
\renewcommand{\theequation}{S\arabic{equation}}

\maketitle
\thispagestyle{empty}
\noindent\textbf{Introduction}

In this document, we first discuss details needed to reproduce the model results and derive equations stated in the main text and then provide some additional results and discussions. This document is organized as follows. In section \ref{sec:model}, we describe the numerical implementation of the hole model; in section \ref{sec:assumptions}, we discuss model assumptions; in section \ref{sec:topos}, we describe different models used to generate synthetic topographies; in section \ref{sec:compare}, we describe the relationship between the parameters in Eqs. 12 and 14 of the main text and parameters of the full 2d model; in section \ref{sec:data}, we describe the measurements used to estimate model parameters; in section \ref{sec:g}, we derive the universal function, $g(\eta)$; in section \ref{sec:heat}, we derive the hole opening timescale, $T_\text{h}$; in section \ref{sec:params}, we show how the minimum pond coverage, $p_\text{min}$, depends on the measurable properties of sea ice; finally, in section \ref{sec:holedist}, we discuss the hole opening distribution, $F$, and show that only its tail affects the pond evolution.
\\
\newpage
\section{Details of the hole model} \label{sec:model}


We implemented our model on a grid, typically $500\times 500$ grid points in size. For each grid point, $\mathbf{x}$, we tracked two fields - ice surface height, $h(\mathbf{x})$, and water level, $w(\mathbf{x})$. Points where $w(\mathbf{x})  > h(\mathbf{x})$ were defined as ponds. We assumed that each connected pond has the same water level, but different ponds can have different water levels. For convenience, we defined $w$ to be equal to ice height for non-ponded grid points, $w(\mathbf{x}) = h(\mathbf{x})$. We define the sea level to be the origin of the vertical axis, so that $w(\mathbf{x}) = 0$ signifies that ponds are at sea level. We initialized the model by generating a random surface, $h(\mathbf{x})$, according to a topographic model (see section \ref{sec:topos}), and setting an initial water level. Typically, we completely covered the surface with water by prescribing a water level equal to the maximum ice height. In this way, the pond coverage fraction at the beginning of a typical simulation was equal to 1. We also prescribed an initial ice thickness, $H$, assumed to be uniform in our model, and shifted $w$ and $h$ by an equal and constant amount to enforce hydrostatic balance. For a water level field as defined above, with $w(\mathbf{x}) = h(\mathbf{x})$ for non-ponded regions, the hydrostatic balance constraint is simply
\begin{linenomath*}
\begin{equation} \label{hyd_bal}
\langle w(\mathbf{x}) \rangle = \frac{\rho_w-\rho_i}{\rho_w} H \text{ ,  }
\end{equation}
\end{linenomath*}
where $\langle ... \rangle$ stands for the average over all grid points, and $\rho_w$ and $\rho_i$ are water and ice densities. Before running the model, we scaled the ice surface height field to have a standard deviation of at most 2\% of ice thickness, $H$. This ensured that suddenly removing all the water from the ice surface would leave no depressions below sea level after hydrostatic balance is enforced. 

We considered each grid point to be a potential hole, and to each one, we ascribed a ``critical temperature,'' $\theta_c(\mathbf{x})$, above which it opens. These critical temperatures were independently drawn from a prescribed probability distribution, $f_\theta$. As a default, we used a normal $f_\theta$. The mean and variance of this distribution are arbitrary, so we set the mean to 0 and variance to 1. We kept track of the ``bulk ice temperature,'' $\theta$. All the grid points that had a critical temperature below the current bulk ice temperature were considered open. We set the initial ice temperature to be below the critical temperature of all grid points. We note that the ``critical temperature,'' $\theta_c(\mathbf{x})$ and the ``bulk ice temperature,'' $\theta$, are simply a way to diagnose where the holes open in our model and need not be related to actual ice temperature.

A model step consisted of several sub-steps:
\begin{enumerate}

\item First, we increase the bulk ice temperature, $\theta$, by a fixed amount, $\text{d} \theta$, and open holes at locations where the new bulk temperature exceeds the critical temperature. Multiple holes can open at each step. The amount, $\text{d} \theta$, by which temperature increases within a given interval of time, $\text{d} t$, depends on the pre-defined hole opening timescale, $T_\text{h}$. This timescale is defined such that the bulk temperature increases by one standard deviation of the prescribed critical temperature distribution, $f_\theta$, during $T_\text{h}$. Since we set the standard deviation of $f_\theta$ to 1, this means that we choose $\text{d} \theta = \frac{\text{d} t}{T_\text{h}}$. 

\item Next, we drain the ponds through these newly opened holes. We simulate the drainage iteratively. We first check which holes are active, i.e., are ponded ($w(\mathbf{x}) >  h(\mathbf{x})$) and are above sea level ($w(\mathbf{x}) >  0$). Next, we find the ponds connected to these active holes. Then, we decrease the water level for each of these ponds by a small amount. This changes the pond coverage slightly, and we find the new ponds as regions with $w(\mathbf{x}) > h(\mathbf{x})$, and enforce $w(\mathbf{x}) = h(\mathbf{x})$ otherwise. Again, we identify holes that are still active, find ponds connected to these holes, and repeat the drainage process until all the holes become inactive. A hole becomes inactive when either a hole's pond water level reaches the ice height at the location of the hole, $w(\mathbf{x}) = h(\mathbf{x})$, or when pond water level reaches sea level, $w(\mathbf{x}) = 0$. 

\item Once drainage through all of the newly opened holes finishes, we preferentially melt ponded ice. We do this by decreasing ice surface height, $h$, in ponded regions while leaving it unchanged in bare ice regions. The amount by which the ponded ice melts depends on the prescribed melt rate, $\text{d} h_\text{diff} / \text{d} t$. We can neglect the melting of bare ice since it is only the relative melt that changes the topography and affects pond evolution. This is justified because we treat ice thinning independently from surface melt, by prescribing a separate parameter, $\text{d}H /\text{d}t$, that we can change irrespective of $\text{d} h_\text{diff} / \text{d} t$, as we explain below. 

\item Finally, we enforce Eq. \ref{hyd_bal} to maintain hydrostatic balance. We do this by shifting the entire ice surface by some amount. Water level is also shifted by the same amount except for ponds at sea level that contain a hole and are maintained at sea level. By returning ice to hydrostatic balance, we effectively adjust the sea level relative to mean ice height.

\end{enumerate}
After adjusting the hydrostatic balance a model step is complete. Then the whole process repeats until all the ponds are at sea level. In the original version of the model, we neglected ice thinning, but when studying stage III of pond evolution, we also thinned the ice using a prescribed rate, $\text{d}H /\text{d}t$, before enforcing hydrostatic balance. In the simulations that include ice thinning, we treat the ice thinning rate as an independent parameter from the surface melt rate, so that $\text{d} h_\text{diff} / \text{d} t$ controls topography evolution while $\text{d}H /\text{d}t$ controls ice thickness evolution, and we do not prescribe pond and bare ice melt rates individually. We also tested versions of the model where not every grid point contains a hole, and where more than one hole can form at each grid point. These modifications did not affect our conclusions in any noticeable way. 

In summary, Table \ref{table:2} shows variables and parameters that enter our model.
\begin{table}[!ht]\label{table:2} 
\centering
\caption{A table of variables and parameters of the hole model.}
\begin{tabular}{| c | p{7cm} |} 
\hline
 Variable	& Name \\
\hline
$\mathbf{x}$  & Coordinate  \\
$h(\mathbf{x})$ & Ice height above sea level \\
$w(\mathbf{x})$ & Water level \\
$\theta$ & Bulk ice temperature\\
$\theta_c(\mathbf{x})$ & Critical temperature of individual brine channels \\
$f_\theta$ & Critical temperature distribution \\
$\text{d}h_\text{diff}/\text{d}t$ & Melt rate of ponded ice relative to bare ice  \\
$T_\text{h}$ & Hole opening timescale \\
$H$ & Ice thickness   \\
$\text{d}H/\text{d}t$ & Ice thinning rate \\
\hline
\end{tabular}
\end{table}

\newpage
\section{Model assumptions}\label{sec:assumptions}

Our model was aimed at explaining the observation that ponds organize around the percolation threshold and exploring the implications of that observation. As such, the model contained only the necessary elements to explain the organization around the percolation threshold, and we have made several simplifying assumptions. All of these assumptions are likely well-justified for undeformed first-year ice, and our model seems to be a good first-order representation of pond evolution after the beginning of drainage. However, in some cases the assumptions we made may not hold. We now discuss violations of our assumptions and how some of them might be accounted for in a future version of the model.

\begin{enumerate}

\item First, by separating pond drainage from the melting and hole opening steps, we have effectively assumed that drainage happens instantaneously. This assumption is supported by observations that find that large ponds can be drained in a matter of hours \citep{polashenski2012mechanisms}, whereas melting through the thickness of the freeboard occurs on the timescale of several days. One reason why this assumption may not hold is that it takes time for brine channels to enlarge to macroscopic sizes. Although there are no measurements of the hole enlargement process, a simple model by \citet{polashenski2012mechanisms} suggests that this may occur on timescales of roughly a day, comparable to the timescale of melt. Another reason is that ponds are partially replenished by additional meltwater flowing into them. The rate of drainage depends primarily on the hole radius, the hydraulic head and the number of holes draining a particular pond, while the rate of replenishing scales with the ice melt rate and the area of the pond. So, large ponds replenish water more easily and it may be necessary to wait until several holes open and grow to a certain size before such ponds can actually start draining. For this reason, our assumption of quick drainage may be invalid especially at the beginning of stage II when ponds cover a large area. We tested our model including a non-negligible time to drain ponds, $T_\text{d}$. In this case we did not separate the drainage and melting steps, but rather melted the ice and adjusted the hydrostatic balance after each increment of drainage, and we assumed that each pond requires $T_\text{d}$ time to drain. Our simulations included drainage timescales of up to 60\% of the hole opening timescale, $T_\text{h}$, and up to 30\% of the memorization timescale, $T_\text{m}$. We found that simulations with long drainage timescales deviated somewhat from the universal function, $g(\eta)$, with the post-drainage pond coverage deviating up to $10\%$ from the instantaneous drainage predictions. Therefore, although this effect is likely not of primary importance, it may not be negligible during the initial drainage period. 

\item Second, we assumed that ice is very flat at the beginning of stage II. The topography of undeformed first-year ice that underlies the snow cover at the beginning of stage I is likely very flat. However, it is questionable whether first year ice at the beginning of stage II can also be considered to be flat since different rates of melt of ponded ice, bare ice, and snow during stage I may act to amplify topography variations. Non-negligible variations in the topography will change the time it takes for different regions of ice to fall below sea level. For this reason, it may be necessary to include the effects of non-negligible initial topography variance in order to get accurate estimates of pond evolution.

\item Third, we assumed that none of the physical parameters change with time, which may not be correct. In particular, the two timescales, $T_\text{m}$, and, $T_\text{h}$, both depend on parameters such as the solar flux or ice thickness that are not constant. Furthermore, the warming rate of the ice interior that enters the hole opening timescale depends on the temperature of the ice, and so changes as the ice warms. Short-time fluctuations, such as variation due to daily changes in solar flux, likely do not contribute significantly to pond evolution. However, longer time variation in these parameters likely affect pond evolution quantitatively, although qualitative conclusions likely remain the same. Time variability can be simply added to our model. In this case, instead of using Eqs. \ref{eqn:12} and \ref{eqn:14} of the main text, pond coverage evolution would be found by solving a differential equation 
\begin{linenomath*}
\begin{equation}\label{pitime}
\frac{\text{d} \Pi}{\text{d} t} =  \frac{\text{d} g}{\text{d} \eta} \Big(\frac{\partial \eta}{\partial t} + \sum_i \frac{\partial \eta}{\partial P_i}\frac{\partial P_i}{\partial t} \Big) \text{  , }
\end{equation}
\end{linenomath*}
where $P_i$ stand for all the parameters that change with time. The term $\text{d} g/ \text{d} \eta$ is the derivative of the universal function with respect to its argument and is approximately given by Eq. \ref{eqn:4} of the main text.

\item Next, we assumed that water flowing towards the holes does not change the ice topography. Faster flowing water will exchange heat with the underlying ice at a higher rate. This will then lead to the formation of channels near the drainage holes that focus water and enhance melting. These channels have been observed in the field near large holes  \citep{polashenski2012mechanisms,landy2014surface}. They have the potential to impact the pond evolution by allowing a single hole to drain larger portions of the ice. To assess the effect of channel formation, we added a crude channel formation scheme to our model. To allow for channel formation, we increased the melt rate along the medial line of ponds while they were draining. To include some elements of realism, we made this increase in melt rate inversely proportional to the distance from the hole, inversely proportional to the local width of the pond in the transverse direction of the medial line, and proportional to the volume of water flowing through the hole, assuming that the water volume is equally partitioned between each hole that drains a particular pond. Such a setup led to the formation of channels between ponds with the most intense channeling occurring during initial drainage since the highest water volume is drained at that time. These assumptions are not highly realistic but we believe they are sufficient for this initial test. We ran several such simulations changing the intensity of melt along the medial line. We found that for a high enough channeling rate, a single hole may be enough to drain the entire surface. However, for any intermediate value, we found that the curve $g(\eta)$ remains approximately the same, and the effect of channel formation can be summarized by simply adjusting the values of $p_c$ and $c$. In particular, the highest channeling rate we tested that did not immediately drain the whole surface reduced $p_c$ on a ``diffusion'' topography from $p_c = 0.5$ to $p_c \approx 0.27$ and $c$ from $c\approx 4.1$ to $c\approx 1.4$. This suggests that Eqs. \ref{eqn:12} and \ref{eqn:14} of the main text may still be used in the presence of channel formation, but the values of $p_c$ and $c$ may depend on channel formation physics. The fact that the effect of channel formation can be summarized by simply adjusting the values of $p_c$ and $c$ in our scheme is likely due to the fact that most channeling occurs after drainage through the first several holes, so the topography remains relatively unchanged as more holes open. This likely also occurs in the field, but in a scenario where the surface is modified throughout pond drainage, the values of $p_c$ and $c$ may need to be made time-dependent as well. Exploring this hypothesis is beyond the scope of this paper. We note that \citet{popovic2018simple} estimated that the percolation threshold for late-summer ponds photographed during the 1998 SHEBA mission was around 0.3, while it was around 0.4 for ponds photographed during the 2005 HOTRAX mission. This is in contrast with the ``snow dune'' topography which accurately describes the pre-melt surface conditions on first-year ice \citep{popovic2020snow} and predicts a percolation threshold of roughly between 0.4 and 0.5, with values between 0.45 and 0.5 being more likely based on the LiDAR measurements of the pre-melt snow topography \citep{popovic2020snow}. Therefore, it is possible that channel formation by water flow decreases the percolation threshold of the pre-melt snow topography by an amount on the order of 0.1 to 0.2.

\item We also assumed that at the beginning of stage II, ice is completely covered by water. This assumption is unrealistic as ice is typically covered significantly less at peak coverage. However, this assumption does not change any of the conclusions we made about pond evolution. As long as the pond coverage fraction at the beginning of the drainage stage is above the percolation threshold, pond evolution will progress identically after drainage below the percolation threshold following the opening of the first several holes. The only effect of this assumption is that non-uniform melt rates before the beginning of drainage may amplify the topography variations as already discussed in item 2 above.   

\item Next, we assumed that drainage happens entirely through large holes and not through the bulk of the ice. This is in contrast with all previous melt pond models that modeled melt ponds as a balance between meltwater production and drainage through the bulk of the ice \citep[see, e.g., ][]{luthje2006modeling,skyllingstad2015simulation}. \citet{polashenski2012mechanisms} observed that during stage II, water is almost entirely drained through large holes. Furthermore, \citet{eicken2004hydraulic} found that early in the melt season, ice permeability is negligible. However, \citet{polashenski2017percolation} found that very large brine channels may not close fully, and there may be some leftover bulk permeability of the ice. A significant amount of drainage through the bulk of the ice during stage II could in principle qualitatively alter the conclusions of our model. In particular, since the percolation threshold does not control bulk drainage, $p_c$ would no longer be a significant coverage fraction and the universal curve $g(\eta)$ would no longer control pond evolution. However, since direct observations by \citet{popovic2018simple} show that ponds seem to be organized around the percolation threshold, bulk ice drainage likely has only a small effect.  

\item We assumed that ice is a rigid plate that cannot elastically flex as the ponds drain. This implies that hydrostatic balance is a global condition that determines the height of the freeboard, rather than being determined by local mass-balance. Equation \ref{eqn:5} of the main text follows from this assumption. This assumption is likely well-justified as ice flexure following pond drainage has not been observed in field. The rigid-plate approximation is also employed in all current melt pond models. 

\item When simulating stage III, we assumed that ponds grow entirely due to ice thinning and we have neglected lateral melting. This assumption seems justified at least on first-year ice, as observations show that freeboard sinking due to ice thinning is the dominant mode of pond evolution during stage III \citep{polashenski2012mechanisms,landy2014surface}, although some models suggest lateral melting may be important \citep{scagliarini2018modelling}. Lateral melt would increase pond coverage during stage III, and could be included in our model by adding a lateral melt contribution as described in \citet{popovic2017simple}.

\item We assumed that ice represents a rigid barrier for water flowing horizontally. This is generally true, although water may be able to flow between disconnected ponds if they are only separated by a narrow strip of ice. This was demonstrated by \citet{eicken2002tracer} who showed that tracers released in disconnected ponds can actually get mixed. However, this effect is most likely small. We note that while ice is mainly impermeable to horizontal flow of water, snow is not, so our ``snow dune'' topography is actually meant to represent ice topography left after snow has melted away.  
\end{enumerate}

\newpage
\section{Synthetic topographies} \label{sec:topos}

To test our model, we used several types of synthetic topographies. Each of these topographies had different statistical characteristics such as the surface height distribution and the percolation threshold which allowed us to test the generality of our results. Importantly, each of these topographies was well-described by a single characteristic length-scale, which is necessary in order for the universal function $g$ to describe the drainage well.

\begin{figure*}
\centering
\includegraphics[width=1.\linewidth]{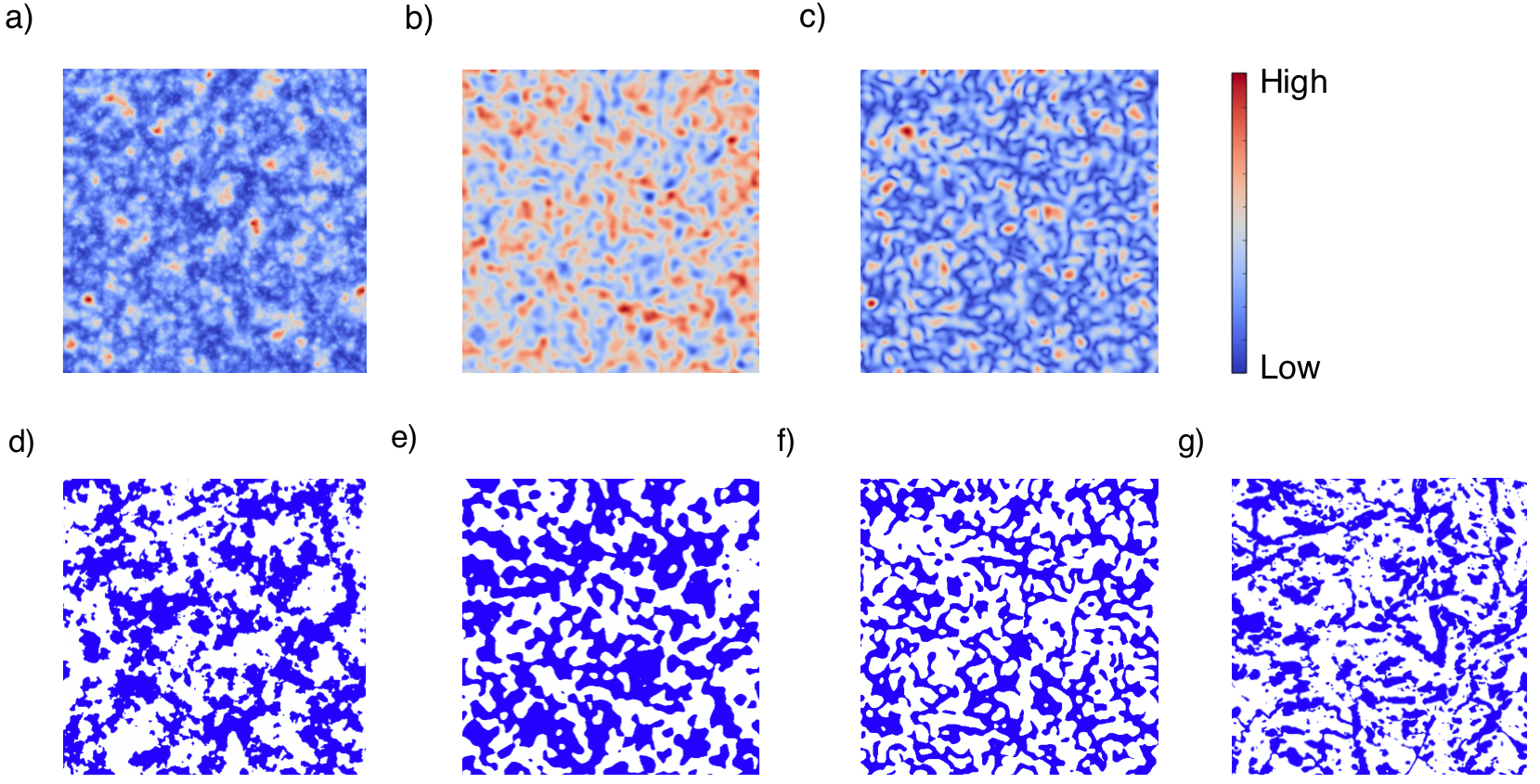}
\caption{a-c: Examples of different types of topographies we used. Red colors indicate highs of the topography, while blue colors indicate lows of the topography. a) A ``snow dune'' topography. b) A ``diffusion'' topography. c) A ``Rayleigh'' topography. d-f: Examples of ponds on different topographies. d) Ponds on a ``snow dune'' topography. e) Ponds on a ``diffusion'' topography. f) Ponds on a ``Rayleigh'' topography. g) A binarized image of a real melt pond photograph taken on August 14th during the HOTRAX mission.}
\label{fig:topos}
\end{figure*}

Our default was the ``snow dune'' topography (Fig. \ref{fig:topos}a). Ponds that form on this surface are shown in Fig. \ref{fig:topos}d. This topography is a generalization of the void model described in \citet{popovic2018simple}, and ponds that form on this topography therefore reproduce the pond geometry well. In \citet{popovic2020snow}, we also showed that this topography reproduces the measured snow topography on undeformed ice highly accurately. We note that even though this topography was meant to describe the properties of the pre-melt snow surface, here we will consider it as a representation of impermeable ice. We generate this topography as a sum of $N_\text{m}$ mounds placed randomly on an initially flat surface. These mounds have a Gaussian shape, $h(\mathbf{x}) = h_m e^{-\frac{(\mathbf{x}-\mathbf{x}_0)^2}{2 r^2}}$, and a horizontal scale, $r$, randomly chosen from an exponential probability distribution, $f_r$, with a typical scale $r_0$ , $f_r(r) = \frac{1}{r_0}e^{-r/r_0}$, equivalent to the distribution of circle radii in the void model of \citet{popovic2018simple}. To prevent having unrealistically narrow and high mounds, we prescribed the height of each mound, $h_m$, to be proportional to its horizontal scale. Optionally, the mounds may also be elongated along a certain axis and the axes may be preferentially aligned to simulate anisotropy in the snow dunes. Including anisotropy did not change any of the conclusions of our model. The ``snow dune'' surface has a height distribution that is well-fit with a gamma distribution (see \citet{popovic2020snow} for more details). Parameters of the surface height distribution depend on the density of mounds placed, $\rho \equiv N_\text{m} \frac{r_0^2}{L^2}$, where $N_\text{m}$ is the number of mounds placed within the domain of size $L$. When few mounds are placed, such that $\rho \ll 1$, the height distribution is highly right-skewed, while when many mounds are placed, such that $\rho \gg 1$, the height distribution converges to a Gaussian. For this reason the percolation threshold also depends on the density of mounds. Specifically, when $\rho$ is large, $p_c = 0.5$ due to the symmetric height distribution. When the height distribution significantly deviates from a Gaussian, $p_c < 0.5$. In \citet{popovic2020snow}, we find that the ``snow dune'' topography with approximately $0.2 < \rho < 0.5$ reproduces the LiDAR measurements of snow on first-year ice. Using $\rho$ in this range, we find that the percolation threshold of a realistic ``snow dune'' topography lies approximately between 0.4 and 0.5. 

In addition to the ``snow dune'' topography, we also used two other types of topographies. The first type we called the ``diffusion'' topography (Fig. \ref{fig:topos}b). Ponds that form on this surface are shown in Fig. \ref{fig:topos}e. This topography is generated by first independently assigning a height to each grid point at random, and then letting this configuration diffuse for some time. Thus, to generate this topography, we numerically solve the diffusion equation. Diffusion smooths out variability in height up to a certain length scale that depends on the time allowed for diffusion. This length scale then determines the length scale $l_0$. To obtain the desired $l_0$, we set the diffusion coefficient to 1 and tune the time for diffusion. This surface has a Gaussian height distribution. Due to this fact, the percolation threshold is $p_c = 0.5$. We note that a similar type of topography has been previously used to model melt pond geometry \citep{bowen2018modeling}. 

We called the other type of topography we used to test our model the ``Rayleigh'' topography (Fig. \ref{fig:topos}c). Ponds that form on this surface are shown in Fig. \ref{fig:topos}f. This topography is made by generating two ``diffusion'' topographies, $h_1(\mathbf{x})$ and $h_2(\mathbf{x})$, with the same length-scale but initialized with a different random configuration, and then taking the square root of the sum of their squares, $h(\mathbf{x}) = \sqrt{h_1^2(\mathbf{x}) + h_2^2(\mathbf{x})}$. This surface has a non-symmetric Rayleigh height distribution (hence the name). The percolation threshold on this surface is $p_c \approx 0.4$. A percolation threshold that deviates from 0.5 is the main reason we tested this surface. ``Diffusion'' and ``Rayleigh'' surfaces generated as above have no obvious relation to any physical feature in sea ice, but are useful in order to test the universality of drainage.  

When generating each of the topographies above, for convenience, we assumed periodic boundary conditions. To estimate $p_c$ for each synthetic topography, we cut the topography with a horizontal plane and shifted it up or down, tracking the connected clusters of the surface that lay below this plane. We found a position of the plane at which a cluster that spans the domain first appears, and estimated $p_c$ as the fraction of the surface below the plane at this level. A constant $c$ in the parameter $\eta = cNl_0^2/L^2$ is a property of a topography type in a similar way as $p_c$. In section \ref{sec:g}, we show that this constant can be related to other properties of the surface such as the percolation threshold and the amplitude of the cluster correlation length near the percolation threshold. However, here we determined $c$ empirically by finding a value for which curves $\Pi = g(\eta)$ on synthetic topographies best fit  $g(\eta)$ estimated analytically using Eq. \ref{eqn:4} of the main text according to the least-squares metric. Similar to $p_c$, a constant $c$ for ``snow dune'' topographies could in principle depend on the density of mounds, $\rho$. However, we found that $c \approx 3$ for all tested densities of mounds. We found that for the ``diffusion'' surface $c \approx 4.1$ and $c \approx 3$ for the ``Rayleigh'' surface.

\newpage
\section{Comparing the full 2d model to the estimate} \label{sec:compare}

In Fig. \ref{fig:3} of the main text, we compared the full 2d hole model to Eqs. \ref{eqn:12} to \ref{eqn:14} of the main text. To make this comparison, we needed to specify the hole opening distribution, $F$, the universal function $g$, and the parameters $T_\text{h}$, $T_\text{m}$, $t_0$, $l_0$, $L$, $N_0$, $c$ and $p_c$ in Eqs. \ref{eqn:12} to \ref{eqn:14} of the main text. Here, we describe in detail how we related the parameters inputed in the full 2d model to parameters that enter Eqs. \ref{eqn:12} to \ref{eqn:14} of the main text. Note that the parameters of the 2d hole model are not related to Eqs. \ref{eqn:12} to \ref{eqn:14} via Eqs. \ref{eqn:5} to \ref{eqn:11} of the main text that describe how measurable physical parameters enter the pond evolution equations since not all of these physical parameters are prescribed in the 2d model (see Table \ref{table:2} for the list of parameters directly prescribed in the 2d model).

\begin{enumerate}
\item We derived the hole opening distribution, $F$, as the cumulative distribution of the distribution of critical temperatures, $f_\theta$, defined in the full 2d model. This relationship is correct because we used a constant rate of bulk ice temperature increase in the 2d model. Since, as a default, we used a normal $f_\theta$, the default $F$ was a cumulative normal distribution. 

The universal function, $g$, is simply the solution to Eq. \ref{eqn:4} of the main text. 


\item The hole opening timescale, $T_\text{h}$, was directly prescribed in the full 2d model as the time for the bulk ice temperature to increase by one standard deviation of the distribution $f_\theta$. Therefore, we used this timescale in Eqs. \ref{eqn:12} to \ref{eqn:14} of the main text as well. 

\item The memorization timescale, $T_\text{m}$, was not prescribed directly in the 2d model, so we had to derive it in order to use it in Eqs. \ref{eqn:12} to \ref{eqn:14} of the main text. To find $T_\text{m}$ in simulations with $\text{d}H/\text{d}t = 0$, we used ice thickness, $H$, and the preferential ponded ice melt rate, $\text{d} h_\text{diff} / \text{d} t$, prescribed in the full 2d model to simultaneously solve for $p_\text{min}$ and $T_\text{m}$ using Eqs. \ref{eqn:7} and \ref{eqn:13} of the main text. 

\item The center of the hole distribution, $t_0$, fluctuates slightly for each run of the 2d model. For this reason in Eqs. \ref{eqn:12} to \ref{eqn:14} of the main text, we did not use the approximate relation for $t_0$ defined in Eq. \ref{eqn:9} of the main text. Instead, we explicitly recorded the timing of the center of the hole distribution relative to the timing of the first hole for each run of the 2d model and used that number in the corresponding estimate. Using Eq. \ref{eqn:9} of the main text to find $t_0$ also gives relatively good results, but there are noticeable deviations between the estimate and the actual model solution. 

\item For the typical pond size, $l_0$, we used the length at which the autocorrelation function for a configuration of ponds in the 2d model drops by a factor of $e$. 

\item Domain size, $L$, is simply the number of grid points on the side of the domain in the full 2d model. 

\item Potential number of holes, $N_0$, in simulations where each grid point represents a potential hole was equal to the total number of grid points, $N_0 = L^2$. In simulations where there can be more than one hole per grid point, $N_0$ is correspondingly larger. 

\item Parameters $c$ and $p_c$ are properties of the topography type, so in the estimates, we used the parameters of the corresponding topographies we used in the full 2d simulations. As we described in section \ref{sec:topos}, for each topography type, we determined $c$ by least-squares fitting the drainage curves to Eq. \ref{eqn:4} of the main text, and we estimated $p_c$ as the fraction of the surface below a level plane that cuts through the topography such that a connected level set that spans the domain first appears. 

\end{enumerate}
Relating the parameters as described above, we were able to uniquely compare estimates using Eqs. \ref{eqn:12} and \ref{eqn:13} of the main text to the full 2d hole model for stage II in the absence of ice thinning. This comparison was shown in Fig. \ref{fig:3}a of the main text. In Fig. \ref{fig:3}b of the main text, we included ice thinning in the full 2d model to simulate stage III. There, we prescribed a constant ice thinning rate, $\text{d}H/\text{d}t$, and the initial ice thickness, $H_0$, such that ice thickness evolves as
\begin{linenomath*}
\begin{equation}\label{thin}
H(t) = H_0 - \frac{\text{d}H}{\text{d}t}t \text{  . }
\end{equation}
\end{linenomath*}
When making the estimate with ice thinning included, most of the parameters that enter Eqs. \ref{eqn:12} to \ref{eqn:14} of the main text can be related to the 2d model parameters in the same way as described above, with the most notable difference being the memorization timescale, $T_\text{m}$. Equation \ref{eqn:7} of the main text defines $T_\text{m}$, as the time for ponded ice to melt below sea level in terms of ice thickness and pond coverage. Since $T_\text{m}$ is the time it takes for ponded ice to melt through the thickness of \textit{post-drainage} freeboard, the thickness and the pond coverage that enter Eq. \ref{eqn:7} of the main text represent the post-drainage parameters. Therefore, to get $T_\text{m}$, we need to solve Eqs. \ref{eqn:7} and \ref{eqn:13} of the main text, and Eq. \ref{thin} simultaneously for post-drainage coverage, $p_\text{min}$, memorization timescale $T_\text{m}$, and post-drainage thickness, $H(T_\text{m})$. 

In addition to predicting the time at which pond bottoms fall below sea level, $T_\text{m}$ estimated as above also coincides fairly accurately with the time at which the water level throughout the domain reaches sea level. This is because, if pond bottoms lay below sea level, ponds do not become disconnected as the drainage progresses, so a single hole can drain an entire pond, making the drainage highly efficient after $T_\text{m}$. Note that this is only true because, in our model, ice at the beginning of stage II is flat compared to the freeboard thickness, so that pond bottoms are also flat and all of ponded ice falls below sea level at roughly the same time. If this were not the case, it could happen that drainage through a hole splits a pond into disconnected parts thereby leaving a part undrained, even if the pond bottom was initially below sea level. The fact that drainage through holes becomes efficient once pond bottoms fall below sea level means that Eq. \ref{eqn:12} of the main text is approximately valid up to $T_\text{m}$, and that Eq. \ref{eqn:14} of the main text is approximately valid after $T_\text{m}$. $T_\text{m}$ estimated in this way is also a crossover time when $p(t)$ estimated using Eq. \ref{eqn:12} of the main text becomes less than $p(t)$ estimated using Eq. \ref{eqn:14} of the main text. Like Eq. \ref{eqn:13}, Eq. \ref{eqn:14} of the main text cannot be solved directly since there the pond coverage $p(t)$ depends on $T_\text{m}(t)$, which in turn depends on $p(t)$ through Eq. \ref{eqn:15} of the main text. Therefore, to solve for pond coverage evolution during stage III, we have to simultaneously solve for $p(t)$ and $T_\text{m}(t)$ using Eqs. \ref{eqn:14} and \ref{eqn:15} of the main text where $H(t)$ is given by Eq. \ref{thin}.

\newpage
\section{Satellite and field data analysis}\label{sec:data}

In the main text we used field measurements and satellite data to constrain our model and to test its predictions. In this section we describe these measurements and discuss how we used them to estimate parameters that enter our model.

\subsection{Estimating parameters based on field measurements}\label{sec:subField}

\citet{polashenski2012mechanisms} collected extensive field data aimed at understanding the formation and evolution of melt ponds. The experiment was performed on land-fast first-year ice near Barrow, Alaska and was repeated during the summers of 2008, 2009, and 2010, with the most extensive studies done during 2009. Measurements of pond coverage shown in Fig. \ref{fig:4}a of the main text were made along a 200 m-long transect every few days during 2009. Additionally, they made measurements including ice thickness, temperature and salinity, ice and pond albedo, meltwater production and drainage, and ice and snow topography. Based on their measurements and data collected during other studies, we were able to estimate most of the parameters that enter the pond evolution equation. These values are reported as default values in Table \ref{tab:1} of the main text along with a plausible range when it was possible to estimate that.

We estimated the linear size of the drainage basin, $L$, based on the remark of \citet{polashenski2012mechanisms} that the drainage basin where the measurements were made is 1.3 km$\times$1.7 km in size. Therefore, we used $L \approx 1.5\text{ km}$. We estimated $T_\text{m}$ using Eq. \ref{eqn:7} of the main text coupled with their measurements of ice thickness ($H \approx 1.2\text{ m}$), pond and ice albedo ($\alpha_p \approx 0.25$, $\alpha_i \approx 0.6$), and measurements of solar flux from the nearby weather station ($F_\text{sol} \approx 254 \text{ W}\text{ m}^{-2}$ average flux during field experiment). We estimated the total number of potential holes, $N_0$, based on measurements of \citet{golden2001brine} who found that the density of brine channels was between 60 and 120 per $\text{m}^2$ in their measurements. We thus used $N_0 \approx L^2 100\text{ m}^{-2}$. We note that depending on the conditions during ice growth, the density of brine channels may vary substantially \citep{wakatsuchi1985brine}. Nevertheless, as we show in section \ref{sec:params}, pond coverage depends only weakly on brine channel density, so even an order of magnitude difference in brine channel density leads to only several percent change in pond coverage. 

\citet{polashenski2012mechanisms} did not estimate the pond size $l_0$. However, based on \citet{popovic2020snow} the pond size seems to be constant between different years with $l_0 \approx 5.5\text{ m}$. \citet{popovic2018simple} estimated the percolation threshold for late-summer ponds for two different years and found it to be around 0.3 and 0.4 for these years. Our simulations on the ``snow dune'' topography predict $0.4<p_c<0.5$. Based on these estimates and simulations we determine that the percolation threshold likely lays somewhere between 0.3 and 0.5. The constant $c$ had only a small effect in our simulations so we chose $c = 3$, consistent with the snow dune topography. We estimated the center of the hole distribution, $t_0$, based on Eq. \ref{eqn:9} of the main text and we found $t_0 \approx 13.4 \text{ days}$. 

Finally, we estimated $T_\text{h}$ using Eqs. \ref{eqn:10} and \ref{eqn:11} of the main text. This estimate is approximate for two reasons - 1) we are uncertain about the mechanism that drives hole opening and 2) even if the mechanism is correct, some of the parameters that enter Eqs. \ref{eqn:10} and \ref{eqn:11} of the main text are difficult to estimate. In addition to measurements of thickness, albedo and solar flux, we used a measurement by \citet{polashenski2012mechanisms} of ice salinity, $S = 3 \text{ ppt}$. The extinction coefficient, $\kappa$, is relatively well-documented and we used $\kappa = 1.5 \text{ m}^{-1}$ \citep{untersteiner1961mass}. Ice conductivity, $k$, can vary because ice and brine have different conductivities. Using the same conductivity parameterization as in  \citet{polashenski2012mechanisms}, and assuming a brine volume fraction between 0 and 0.5, we found that conductivity can vary between 1.3 and 2$\text{ W}\text{ m}^{-1\circ}\text{C}^{-1}$. We used $k = 1.8\text{ W}\text{ m}^{-1\circ}\text{C}^{-1}$ which corresponds to a brine volume fraction of 0.1. Estimating parameters $\theta_0$, $\Delta \theta$, $c^*$, and $z^*$ is difficult. To estimate the reference temperature, $\theta_0$, and the temperature range, $\Delta \theta$, we used observations made by \citet{polashenski2017percolation}. They noted that holes tend to begin to open when ice interior temperature reaches roughly $-1.6^\circ$C and open completely when ice temperature reaches around $-0.9^\circ$C. Therefore, we chose the temperature range to be $\Delta \theta = 0.7^\circ \text{C}$. Based on these observations and our solutions to the full heat equation, we chose the reference temperature to be $\theta_0 = -1.2^\circ$C, and we used the shape parameter, $c^* = 2$ (see section \ref{sec:heat}). \citet{polashenski2017percolation} found that freshwater plugs do not form at a single depth within the ice, but rather throughout a range of depths. Nevertheless, we used a single $z^*$ for simplicity. \citet{polashenski2017percolation} took photographs that show that freshwater plugs tend not to form in the upper 0.6 m of the ice, and we estimated roughly $z^* = 0.6\text{ m}$ for this reason. Future models will need to include the fact that tracking a single layer within the ice is insufficient to capture the process of hole formation.

\subsection{Comparing the estimated pond coverage to \citet{polashenski2012mechanisms}} \label{sec:subPolash}


With the physical parameters estimated above, we were able to find all of the parameters that enter Eqs. \ref{eqn:12} to \ref{eqn:14} of the main text using Eqs. \ref{eqn:7} and \ref{eqn:9}-\ref{eqn:11} of the main text. We used these solutions to compare our estimate to pond coverage evolution observed by \citet{polashenski2012mechanisms} in Fig \ref{fig:4}a of the main text. Parameters $\theta_0$, $\Delta \theta$, $c^*$, and $z^*$ clearly have larger uncertainty than other parameters that can be directly measured. Therefore, to find the wider error bars in Fig \ref{fig:4}a of the main text (light shaded region), we assumed a $50\%$ error in the parameters $\theta_0$, $\Delta \theta$, $c^*$, and $z^*$, and used a $10\%$ measurement error for all other parameters. For narrower error bars in Fig \ref{fig:4}a of the main text (dark shaded region) we used the $10\%$ measurement error for all parameters. These measurement errors are not representative of real uncertainties, and we chose them mainly for demonstrative purposes. In fact, these errors are likely still underestimated. For this reason, our estimates of pond coverage are uncertain and all we can say is that they are at least consistent with our model.  



\subsection{MODIS satellite data} \label{sec:subMODIS}

A dataset of pond coverage fraction estimates based on satellite measurements is available for download at https://cera-www.dkrz.de/WDCC/ui/cerasearch/. In this dataset, pond fraction is estimated using an artificial neural network that takes as input the reflectances from three channels of the visible spectrum of the MODIS instrument aboard the Earth Observation Satellite TERRA \citep{MODIS_v022}. The observations are available every 8 days, have a spatial resolution of 12.5km, and cover the entire Arctic. To make Fig. \ref{fig:4}b of the main text, we collected estimated pond fractions for all grid-cells at all times during a particular year for which there were measurements available. We only considered measurements that showed a non-zero pond coverage. We then summarized these data as probability distributions for several different years. Due to the relatively coarse resolution of the measurements, the estimated maximum pond coverage is likely somewhat underestimated. 

\newpage
\section{Deriving the universal drainage curve} \label{sec:g}

Here, we will motivate a form of the universal function $g(\eta)$, highlighting the origin of its universality. The goal is to estimate the change in pond coverage fraction after a hole opens at a random location on the ice surface. We are assuming that the first several holes have already driven the ponds to the percolation threshold. 

We introduce the cluster correlation function, $G(l)$, as the probability that, given a randomly chosen point on a pond, another point of distance $l$ away will belong to the same pond. Equivalently, $G(l)$ can be viewed as the average fraction of a circle of radius $l$ that belongs to the same pond as a point chosen randomly on any ponded location. Therefore, a quantity $G(l)2\pi l \text{d}l$ is an average area of a ring of radius $l$ and thickness $\text{d}l$ connected to a hole opened randomly at some ponded location, and an integral of $G$ over the entire 2d domain gives the average area of a pond connected to such a hole. We need to multiply this integral by the pond coverage fraction, $p$, to account for the fact that in our model holes can open anywhere on a surface, not just within ponds. Therefore the mean fraction of the surface connected to a randomly placed hole is 
\begin{linenomath*}
\begin{equation} \label{dphole}
\Delta p = p \frac{2\pi \int_0^\infty G(l)l\text{d}l}{L^2} \text{  . }
\end{equation}
\end{linenomath*}

Equation \ref{dphole} would represent the average change in coverage fraction after opening a hole, if that hole drained the entire pond it was connected to. However, a hole can only drain a fraction of an entire pond. Let us denote that fraction as $f_\text{drain}$. The fraction of the pond drained depends on the relative size of the pond to the autocorrelation length, $l_0$. Namely, when a pond has a size comparable to $l_0$, a hole can drain a significant fraction of that pond. On the other hand, when a pond is very large compared to $l_0$, a small amount of drainage would quickly lead to the formation of disconnected regions, and a hole would be unable to drain a significant fraction of the pond. We can describe a typical linear size of the largest ponds using a ``cluster correlation length,'' $\xi$, defined roughly as the linear extent of the largest connected pond or, equivalently, as the length beyond which $G(l)$ rapidly falls to 0. Therefore, a fraction $f_\text{drain}$ can be viewed as a function of $l_0/\xi$ such that $f_\text{drain}(l_0/\xi) \rightarrow 0$ when $l_0/\xi \rightarrow 0$ and $f_\text{drain}(l_0/\xi) \rightarrow 1$ when $l_0/\xi \rightarrow \infty$. The correlation length, $\xi$, depends on the deviation of pond coverage from the percolation threshold since connected ponds become ever larger as the pond coverage approaches $p_c$. Thus, close to the percolation threshold, ponds are large and $l_0/\xi \ll 1$, so, based on a general Taylor expansion and the above limiting behavior, we have 
\begin{linenomath*}
\begin{equation}\label{fcorrTaylor}
f_\text{drain}(\frac{l_0}{\xi}) \approx a \frac{l_0}{\xi} + \text{...}
	\quad\text{ for }\quad  
\big(1-\frac{p}{p_c}\big) \ll 1 \text{  ,} 
\end{equation}
\end{linenomath*}
where $a$ is some constant. We note that this expansion hides an implicit assumption that $f_\text{drain}(l_0/\xi)$ is analytic in the limit $l_0/\xi \rightarrow 0$ and that $a$ is non-zero. In principle, a form $f_\text{drain}(l_0/\xi) \propto (l_0/\xi)^\alpha$ as $\l_0/\xi \rightarrow 0$ for any $\alpha >0$ would also be possible. So, we can justify the above expansion only a posteriori, by showing that our theoretical prediction matches the full 2d simulations well. Based on the entire discussion above we can now estimate the pond fraction drained after opening a hole as
\begin{linenomath*}
\begin{equation} \label{dpdN}
\frac{\text{d}p}{\text{d}N} = - p \frac{2\pi \int_0^\infty G(l)l\text{d}l}{L^2} f_{\text{drain}}(\frac{l_0}{\xi}) \text{  . }
\end{equation}
\end{linenomath*}
Each term on the right-hand side is a function of pond coverage, so this equation defines a curve of pond coverage as a function of the number of open holes. 

\begin{figure*}[ht]
\centering
\noindent\includegraphics[width=0.9\linewidth]{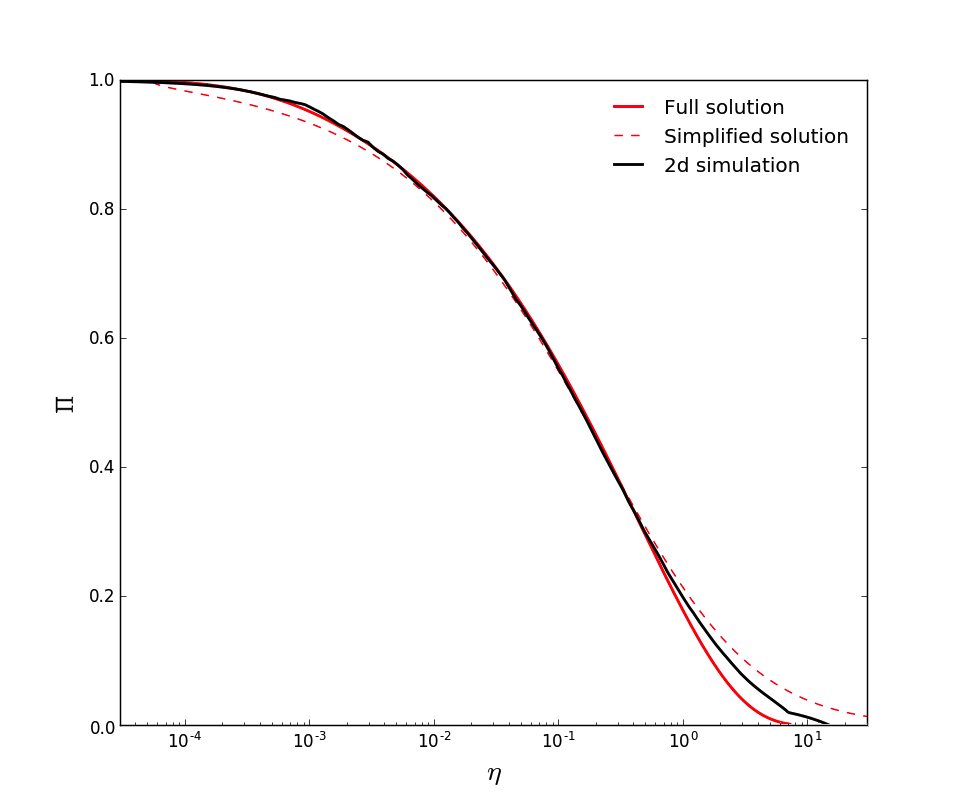}
\caption{a)  Comparing $g(\eta)$ estimated by running a single 2d simulation on a synthetic ``snow dune'' topography with no melt (black line) with a solution of Eq. \ref{dpdN} using the full expressions Eqs. \ref{clustercorr} to \ref{clustercorr_xi_inf} (solid red line) and a solution to universal Eq. \ref{simple} that assumes asymptotic forms of $G(l)$ and $\xi$, Eqs. \ref{clustercorr_theory} and \ref{clustercorr_xi_theory} (dashed red line).}
\label{fig:app1}
\end{figure*}

To close Eq. \ref{dpdN}, we need to estimate the integral $\int_0^\infty G(l)l\text{d}l$ and express $\xi$ in terms of $p$. General percolation theory shows that in the limit of infinite domain size, $L\rightarrow \infty$, close to the percolation threshold, $(1-p/p_c) \ll 1$, and at distances larger than the characteristic scale, $l \gg l_0$, $G(l)$ and $\xi$ behave as
\begin{linenomath*}
\begin{gather} 
G(l) = p(\frac{l}{l_0})^{-5/24}e^{-l/\xi} \label{clustercorr_theory} \text{  , } \\
\xi = A l_0 ( 1 - \frac{p}{p_c} )^{-4/3} \label{clustercorr_xi_theory} \text{  , } 
\end{gather}
\end{linenomath*}
for any model within the percolation universality class \citep{isichenko1992percolation}. Here, $5/24$ and $4/3$ are universal exponents that come from percolation theory, and $A$ is a non-universal order one constant. At distances smaller than $l_0$, $G(l)$ depends on the shape of individual ponds, and, thus, on the details of the topography, but is nevertheless constrained to go to 1 as $l \rightarrow 0$. The facts that 1) both the full $G(l)$ and its asymptotic form, Eq. \ref{clustercorr_theory}, have finite integrals as $l \rightarrow 0$, and 2) that $G(l)$ is significantly above 0 for $l < \xi$, mean that, if $\xi$ is large compared to $l_0$, the contribution of short lengths, $l \ll l_0$, to the integral $\int_0^\infty G(l)l\text{d}l$ can be ignored for both the full $G(l)$ and its asymptotic form, so we can use the asymptotic form, Eq. \ref{clustercorr_theory}, to perform the integration. If the domain size is not infinite, the size of the largest connected ponds is limited by $L$, and, therefore, $\xi$ will not diverge as predicted by Eq. \ref{clustercorr_xi_theory}, but will instead be limited by the domain size. Based on the above discussion, we can use $G(l)$ and $\xi$ given by Eqs. \ref{clustercorr_theory} and \ref{clustercorr_xi_theory} to close the pond drainage equation, Eq. \ref{dpdN}, when $l_0 \ll \xi \ll L$. Moreover, to ensure that we can use Eqs. \ref{clustercorr_theory} and \ref{clustercorr_xi_theory} with Eq. \ref{dpdN} we additionally have to assume that the hole opening process does not disrupt the geometry of the ponds, so that $G$ and $\xi$ have the same form as more and more holes open. Again, the assumption of pond geometry remaining unchanged can only be justified a posteriori, by direct comparison with full 2d simulations. 

We now show that pond drainage is universal in the limit $l_0 \ll \xi \ll L$, i.e., when the pond coverage is neither too close nor too far from the percolation threshold. In that case, we can use the asymptotic forms of $\xi$ and $G(l)$ and the asymptotic expansion Eq. \ref{fcorrTaylor} for $f_\text{drain}$. We can explicitly perform the integral of $G$ over $l$ to get
\begin{linenomath*}
\begin{equation}\label{simple1}
\frac{\text{d} p}{\text{d}N} = - C \frac{l_0^2}{L^2} p^2 (1-\frac{p}{p_c})^{-19/18} \text{  , }
\end{equation}
\end{linenomath*}
where $C = 2\pi \Gamma(\frac{43}{24})aA^{19/24} $ is a numerical constant that depends on $A$, $a$, and constants of integration. Defining $\Pi \equiv \frac{p}{p_c}$ and $\eta \equiv c\frac{l_0^2}{L^2}N$, where $c \equiv p_cC$, we recover Eq. \ref{eqn:4} of the main text
\begin{linenomath*}
\begin{equation} \label{simple}
\frac{\text{d} \Pi}{\text{d} \eta} = - \Pi^2 (1-\Pi)^{-19/18} \text{  . }
\end{equation}
\end{linenomath*}
The appropriate boundary condition for this equation is that ponds are at the percolation threshold when only a finite number of holes are open, $\Pi(\eta = 0) = 1$, where we used the fact that $\eta = 0$ if the domain is infinite and $N$ is finite. We can see that Eq. \ref{simple} defines a universal function $\Pi = g(\eta)$, since there are no non-universal factors that enter it. From the above derivation, we see that the universality of $g(\eta)$ comes from the universality of the percolation theory in the limit $l_0 \ll \xi \ll L$. This derivation also clarifies that the factor $c$ that enters $\eta$ depends on the type of topography because it is a combination of factors $a$, $A$, and $p_c$ which all potentially depend on the details of the topography. A solution to Eq. \ref{simple} is shown with a red dashed line in Fig. \ref{fig:app1}. It agrees well with the simulations apart from at low pond coverage and at pond coverage close to the percolation threshold. Discrepancy between this solution and the simulations near $p_c$ is due to the fact that in this region $\xi \sim L$, and the universality breaks down. We find that this region of disagreement shrinks as we increase our domain size $L$.  


We will now show that we can closely match the simulated $g(\eta)$ also in the regime $p \rightarrow p_c$ if we use $G(l)$ that is valid for all $l$ (including $l < l_0$) and $\xi$ that takes into account the effect of domain size. The matching in this case, however, will not be universal. 

As we have discussed above, Eqs. \ref{clustercorr_theory} and \ref{clustercorr_xi_theory} are asymptotic limits only valid at large separations, $l$ (so that $l \gg l_0$), and when the pond coverage is neither too close nor too far from $p_c$ (so that $l_0 \ll \xi \ll L$). In the Supplementary Information of \citet{popovic2018simple}, we have shown that the $G(l)$ and $\xi$ that fit the measured pond statistics at all lengths, $l$, and finite domain size, $L$, are 
\begin{linenomath*}
\begin{gather} \label{clustercorr}
G(l) = (e^{-l/l_0}(1-p) + p)(1 + \frac{l}{l_0})^{-5/24}e^{-l/\xi} \text{  , } \\
\xi = \xi_{\infty} ( 1 - e^{-BL/\xi_{\infty}} )  \label{clustercorr_xi} \text{  , } \\
\xi_{\infty}  =A l_0 ( 1 - \frac{p}{p_c} )^{-4/3}  \label{clustercorr_xi_inf} \text{  , } 
\end{gather}
\end{linenomath*}
where $\xi_{\infty}$ is the cluster correlation length if the domain size were infinite and has the same form as Eq. \ref{clustercorr_xi_theory}, and $B$ is a non-universal order one constant. These equations respect the limits for $l \gg l_0$ and $L \rightarrow \infty$, and also ensure that $G(l\rightarrow 0)\rightarrow 1$ and that $\xi \propto L$ when $\xi_\infty \rightarrow \infty$. However, the particular form of these functions is arbitrary, and we chose them due to their simplicity and the fact that they match the measured $G(l)$ and $\xi$ highly accurately outside the scaling limit (see SI of \citet{popovic2018simple}). Even though these equations describe $G(l)$ and $\xi$ well for all $l$ and $L$, they only hold close to the percolation threshold, i.e., when $\xi \gg l_0$, so we still expect deviations from them when $p \ll p_c$ (i.e., when $\xi \sim l_0$).

Finally, we show the solution to Eq. \ref{dpdN} using the full Eqs. \ref{clustercorr} to \ref{clustercorr_xi_inf}. To be able to evaluate $\text{d}p/\text{d}N$ in this case, we also need to assume a concrete $f_\text{drain}$ consistent with the asymptotic behavior given in Eq. \ref{fcorrTaylor} and the fact that we require $f_\text{drain}(l_0/\xi) \rightarrow 1$ when $l_0/\xi \rightarrow \infty$. To this end, we choose $f_{\text{drain}} = 1 - e^{-l_0/\xi}$. This arbitrary choice works well because of universality in the scaling limit that we discussed above. With this form of $f_\text{drain}$, we can use Eqs. \ref{clustercorr} to \ref{clustercorr_xi_inf} to solve Eq. \ref{dpdN} and get a function $p = p(N)$. This solution in coordinates $p/p_c$ and $Nl_0^2/L^2$ is shown as a solid red line in Fig. \ref{fig:app1}. We can see that the agreement with the curve arising from a 2d simulation is excellent apart from a small discrepancy at small pond coverage. This discrepancy is due to the breakdown of Eqs. \ref{clustercorr} to \ref{clustercorr_xi_inf} far from $p_c$, due to our ad-hoc choice of $f_\text{drain}$, and due to the fact that opening many holes likely does not leave pond geometry completely unchanged. However, even though we were able to obtain a good match with simulations, this solution is not universal under the rescaling $p \rightarrow \Pi$ and $N \rightarrow \eta$. Namely, changing the non-universal constants $A$ and $B$ or $f_{\text{drain}}$ changes the form of the solution in $\Pi$ and $\eta$ coordinates, so, to get a good match, we had to tune $A$ and $B$ specifically. This again highlights the fact that the universality occurs in the limit $l_0 \ll \xi \ll L$, i.e., when ponds are much larger then the characteristic scale of the topography (set e.g. by the size of snow dunes), and still smaller than the size of the domain (set e.g. by the ice floe size).
\\
\\
\newpage
\section{The heat equation} \label{sec:heat}

In Eqs. \ref{eqn:10} and \ref{eqn:11} of the main text, we noted that the order of magnitude of the hole opening timescale, $T_\text{h}$, can be estimated based on physical properties of ice if we assume that internal temperature at some depth within the ice is the relevant parameter that determines when holes start to open. In this section, following \citet{bitz1999energy}, we will briefly explain how we arrived at this estimate.

Energy supplied to a unit mass of salty ice partly goes into warming the ice and partly into melting the ice around brine pockets to dilute the brine and bring it into equilibrium with the surrounding ice. \citet{ono1967specific} showed that the heat capacity of a unit mass of ice of salinity $S$ and temperature $\theta$ in degrees Celsius can be expressed as
\begin{linenomath*}
\begin{equation} \label{heatcap}
c(\theta,S) = c_0 + \frac{\gamma S}{\theta^2} \text{ ,  }
\end{equation}
\end{linenomath*}
where $c_0 = 2.11 \text{ kJ}\text{ kg}^{-1\circ} \text{C}^{-1}$ is the heat capacity of fresh ice and $\gamma = 18 \text{ kJ}\text{ kg}^{-1} \text{ppt}^{-1\circ}\text{C}$ is a constant. Equation \ref{heatcap} neglects the typically small contribution from the fact that water in brine pockets has a different heat capacity than ice. This equation shows that the heat capacity increases with temperature and diverges when the temperature approaches the melting point of fresh ice. During the summer, ice is in the process of melting, so the ice interior is only several degrees below zero. With typical salinities of 3 ppt or 4 ppt for first year ice, the salinity term in the heat capacity during summer is roughly an order of magnitude larger than $c_0$, so $c(\theta,S) \approx \frac{\gamma S}{\theta^2}$. 

\begin{figure*}
\centering
\noindent\includegraphics[width=1\linewidth]{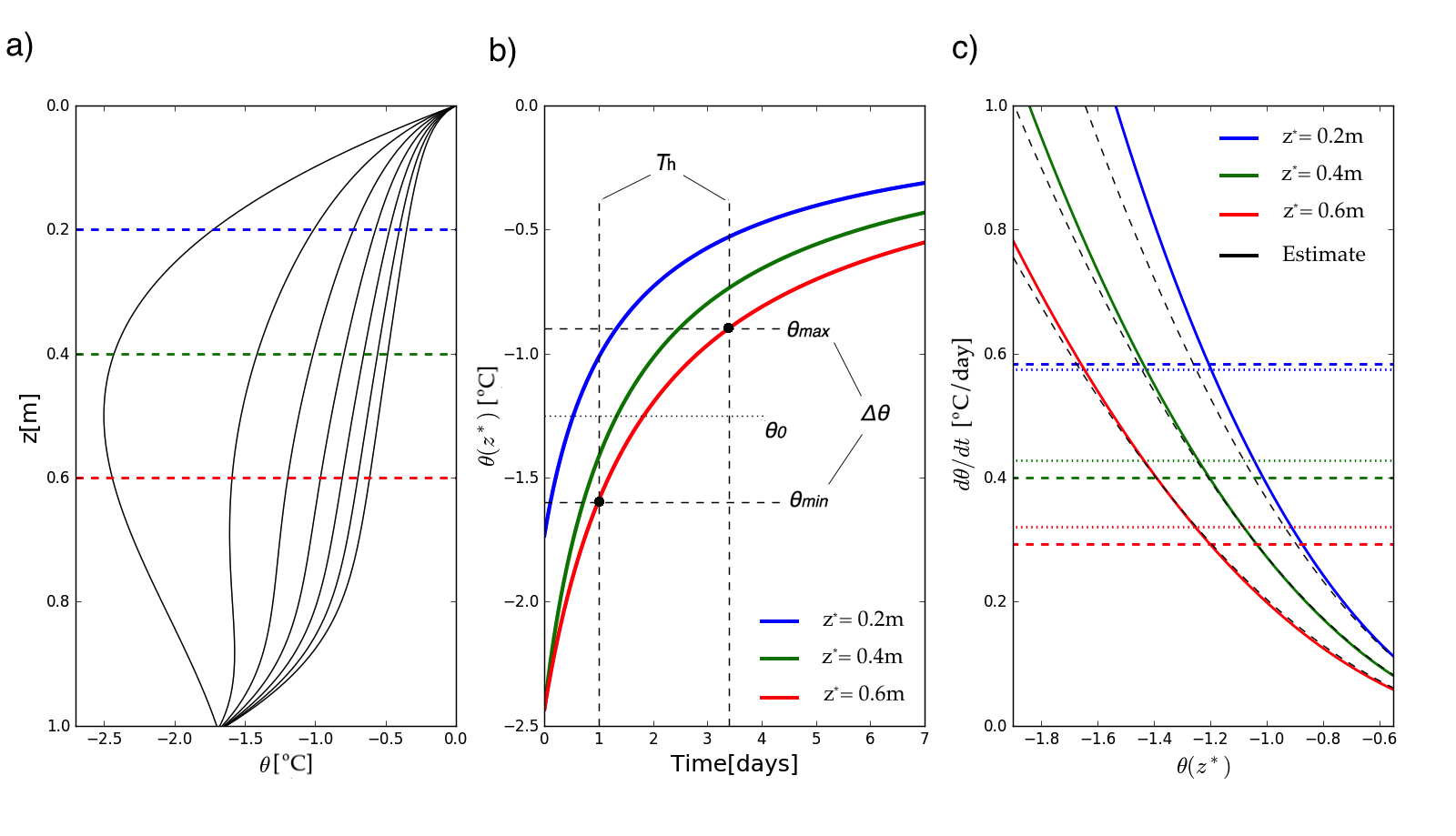}
\caption{ a) Black lines represent represent the vertical temperature profiles within the ice at different times obtained by solving Eq. \ref{partdiff}. Curves to the left correspond to earlier times. Colored horizontal lines represent different choices of $z^*$ that are shown in panels b and c. b) Time evolution of ice temperature at a fixed depth. Different colors stand for different depths, $z^*$, shown in panel a. Horizontal dashed lines represent the temperature at which the holes tend to start opening, $\theta_\text{min}$, and the temperature at which a significant fraction of brine channels are open, $\theta_\text{max}$. The temperature range, $\Delta \theta$, is also marked. Vertical dashed lines represent times at which the temperature at depth $z^* = 0.6 \text{ m}$ crosses $\theta_\text{min}$ and $\theta_\text{max}$, which defines the hole opening timescale, $T_\text{h}$. A reference temperature, $\theta_0$, estimated as the middle of the range between $\theta_\text{min}$ and $\theta_\text{max}$ is also marked with a horizontal dashed line. c) Dependence of the warming rate on current ice temperature. Different colors stand for different depths, $z^*$, marked in panel a. Black dashed curves show estimates using Eq. \ref{heat_est} with $c^* = 2$ and treating $\theta_0$ as a variable. Horizontal solid colored lines show $\text{d}\theta/\text{d}t$ estimated as $\Delta \theta /T_\text{h}$, where $T_\text{h}$ is found numerically as in panel b. Horizontal colored dotted lines show $\text{d}\theta/\text{d}t$ estimated using Eq. \ref{heat_est} with $c^* = 2$ and a reference temperature $\theta_0 = 0.5(\theta_\text{max} + \theta_\text{min})$.}
\label{fig:heat_eq}
\end{figure*}

The ice interior warms because of heat diffusion and because sunlight penetrates beneath the ice surface. This can be expressed with a partial differential equation that determines the warming rate at depth $z$ within the ice
\begin{linenomath*}
\begin{equation} \label{partdiff}
\rho_i c(\theta,S) \frac{\partial \theta}{\partial t}  = \frac{\partial}{\partial z} k \frac{\partial \theta}{\partial z} + \kappa F_\text{0} e^{-\kappa z} \text{ ,  }
\end{equation}
\end{linenomath*}
where $\rho_i$ is the density of ice, $k$ is the thermal conductivity of the ice, $\kappa$ is the extinction coefficient from Beer's law, and $ F_\text{0}$ is the radiative flux that penetrates the upper surface of the ice. The first term on the right hand side represents the contribution from heat diffusion while the second term is the contribution from direct solar heating at a depth $z$. Thermal conductivity, $k$, may in principle depend on depth because brine and ice have different thermal conductivities. This dependence is, however, typically weak. Equation \ref{partdiff} should be supplied with a boundary condition that ice is at its salinity-dependent melting temperature at the top and bottom surface. Furthermore, a temperature at the initial time should also be specified. We solve this equation in Fig. \ref{fig:heat_eq} assuming a uniform salinity profile throughout the ice column and ice bottom temperature fixed at -1.7$^\circ\text{C}$. 

As we have suggested in the main text, our strategy is to estimate $T_\text{h}$ by estimating the warming rate $\frac{\text{d}\theta}{\text{d}t}$ at some depth, $z^*$, within the ice where ice plugs tend to form. In general, the warming rate in Eq. \ref{partdiff} cannot be simply characterized as it depends on the full vertical profiles of temperature and salinity and their history. Nevertheless, we can approximate it as
\begin{linenomath*}
\begin{equation} \label{heat_est}
\frac{\text{d} \theta}{\text{d} t} \approx \frac{\theta_0^2 }{\rho_i \gamma S} \Big( c^* k \frac{\theta_0}{H^2}  + (1-\alpha_p) F_\text{sol} \kappa e^{-\kappa z^*} \Big) \text{  . }
\end{equation}
\end{linenomath*}
Here, we have assumed that the fresh ice heat capacity, $c_0$, is negligible, so that $c(\theta,S) \approx \frac{\gamma S}{\theta^2}$. We have also assumed that thermal conductivity, $k$, does not depend on brine volume fraction, and we have estimated the amount of radiation that penetrates the upper surface as $F_\text{0} \approx (1-\alpha_p) F_\text{sol}$, where $\alpha_p$ is the pond albedo and $ F_\text{sol}$ is the solar radiative flux. We only consider pond albedo because we are only interested in holes that open beneath ponds. A constant $c^*$ accounts for the shape of the temperature profile, while $\theta_0$ and $z^*$ are the reference temperature and depth at which we are estimating the warming rate. 

To be able to estimate the hole opening timescale, $T_\text{h}$, using Eq. \ref{heat_est}, we need to relate the parameters of that equation to physically meaningful quantities. We can define $T_\text{h}$ as the time it takes for ice at a particular depth to warm from a temperature, $\theta_\text{min}$, at which only a small fraction of the brine channels can become holes, to a temperature, $\theta_\text{max}$, at which a significant fraction of the brine channels become holes (see Fig. \ref{fig:heat_eq}b). We denote the difference $\theta_\text{max}-\theta_\text{min}$ as $\Delta \theta$. We thus consider $T_\text{h}$, $\theta_\text{min}$,  $\theta_\text{max}$, and $\Delta \theta$ to be measurable, physically meaningful quantities. We can approximate $\theta_0$ in Eq. \ref{heat_est} to be the middle of the range between $\theta_\text{max}$ and $\theta_\text{min}$, $\theta_0 \approx \frac{\theta_\text{min} + \theta_\text{max}}{2}$. We consider $z^*$ to have physical significance as the depth at which ice plugs form, and can thus be estimated using measurements. Finally, the parameter $c^*$ takes into account the shape of the temperature profile within the ice column. As such, it depends on a multitude of factors such as the boundary conditions, the salinity profile, or the depth at which we are estimating $\frac{\text{d} \theta}{\text{d} t}$. To keep matters simple, we choose $c^* = 2$, corresponding to a quadratic temperature profile with the top and bottom ice temperatures fixed at 0$^\circ\text{C}$. 

To determine how reliable the approximations we made above are, we solved the full heat equation, Eq. \ref{partdiff}, varying the equation parameters. In Fig. \ref{fig:heat_eq}c, we show $\frac{\text{d} \theta}{\text{d} t} (\theta)$ found by solving Eq. \ref{partdiff} numerically, and compare it to $\frac{\text{d} \theta}{\text{d} t} (\theta_0)$ estimated using Eq. \ref{heat_est} treating $\theta_0$ as a variable (compare colored and dashed lines in Fig. \ref{fig:heat_eq}c). We can see that in all cases, the estimate and the numerical simulations behave qualitatively in the same way and are of similar magnitude (within around 20\% of each other). In Fig. \ref{fig:heat_eq}c, we also show $\frac{\Delta \theta}{T_\text{h}}$, where we estimate $T_\text{h}$ by solving Eq. \ref{partdiff} numerically and compare it to $\frac{\text{d} \theta}{\text{d} t}$ estimated using Eq. \ref{heat_est} with $\theta_0 = \frac{\theta_\text{min} + \theta_\text{max}}{2}$. Again, we find that our approximations are of the same order of magnitude as the simulations (in this case within 10\% of each other), which implies that $T_\text{h}$ can be reasonably estimated using Eq. \ref{heat_est}.  

For each particular numerical solution of Eq. \ref{partdiff}, we can improve the match between the simulations and estimates by tuning $c^*$ or changing the definition of $\theta_0$. Depending on the depth, boundary conditions, shape of the salinity profile, and the strength of direct solar heating, we find that $c^*$ can be anything between roughly 1 and 10, while $\theta_0$ can be modified to $\theta_0 = (1-r)\theta_\text{max} + r\theta_\text{min}$, with the numerical factor $r$ anywhere between 0.4 and 0.5. Nevertheless, using $c^* = 2$ and $\theta_0 = \frac{\theta_\text{min} + \theta_\text{max}}{2}$, as we did above, and varying the parameters of the numerical simulation, we find that $\frac{\text{d} \theta}{\text{d} t}(\theta_0)$ estimated using Eq. \ref{heat_est} and treating $\theta_0$ as variable is always of the same order of magnitude as $\frac{\text{d} \theta}{\text{d} t}(\theta)$ estimated using Eq. \ref{partdiff} for all $\theta_0$ and $\theta$, while $\frac{\text{d} \theta}{\text{d} t}$ estimated at $\theta_0 = \frac{\theta_\text{min} + \theta_\text{max}}{2}$ using Eq. \ref{heat_est} is always of the same order of magnitude as $\frac{\Delta \theta}{T_\text{h}}$, with $T_\text{h}$ estimated by solving Eq. \ref{partdiff} numerically. This means that Eq. \ref{heat_est} with $c^* = 2$ and $r = 0.5$ is a reasonable first-order approximation to the full heat equation under any configuration, with errors in estimated $\frac{\text{d} \theta}{\text{d} t}$ likely being on the order of 10-20\% of the full solution based on the results in Fig. \ref{fig:heat_eq}c. 

Finally, using Eq. \ref{heat_est}, we can estimate the hole opening timescale, $T_\text{h}$ as 
\begin{linenomath*}
\begin{equation} \label{Th}
T_\text{h} =\frac{\Delta \theta}{\text{d} \theta / \text{d} t} \text{ , }
\end{equation}
\end{linenomath*}
with $\frac{\text{d} \theta}{\text{d} t}$ estimated using Eq. \ref{heat_est}. In addition to the fact that this estimate may disagree with the full solution to the heat equation as we have discussed above, there are several additional reasons why it is uncertain - 1) it is difficult to precisely define and determine parameters $\theta_\text{max}$, $\theta_\text{min}$, and $z^*$ from measurements, 2) it is unclear whether these parameters themselves depend on physical properties such as ice salinity or thickness, and, most importantly, 3) it is unclear whether the ice interior temperature is in fact a good indicator of when holes start to open. In particular, the assumption that ice interior temperature is a good indicator of hole opening leads to some counter-intuitive conclusions that we discuss in section \ref{sec:params} and section \ref{sec:discussion} of the main text. Namely, it predicts that holes open more slowly when ice is more saline or when it is closer to its melting point, since in these cases $\frac{\text{d} \theta}{\text{d} t}$ is lower due to a higher higher heat capacity stemming from a higher brine volume fraction. This is counter-intuitive because it implies that it is more difficult to create holes when ice is more porous and when more heat is expended into melting the ice pores. For these reasons, improving our understanding of hole formation physics is crucial for understanding melt pond evolution.

\newpage
\section{Dependence of pond coverage on physical parameters}\label{sec:params}

In this section, we will discuss how pond coverage depends on the measurable properties of the ice in our model. In particular, we will focus on the pond coverage minimum, $p_\text{min}$, found using Eqs.  \ref{eqn:7}, \ref{eqn:9}-\ref{eqn:11}, and \ref{eqn:13} of the main text. In the entire discussion below, we will use the cumulative normal hole opening distribution, $F$, and the set of default parameters defined in Table \ref{tab:1} of the main text.  

\begin{figure*}
\centering
\includegraphics[width=1.05\linewidth]{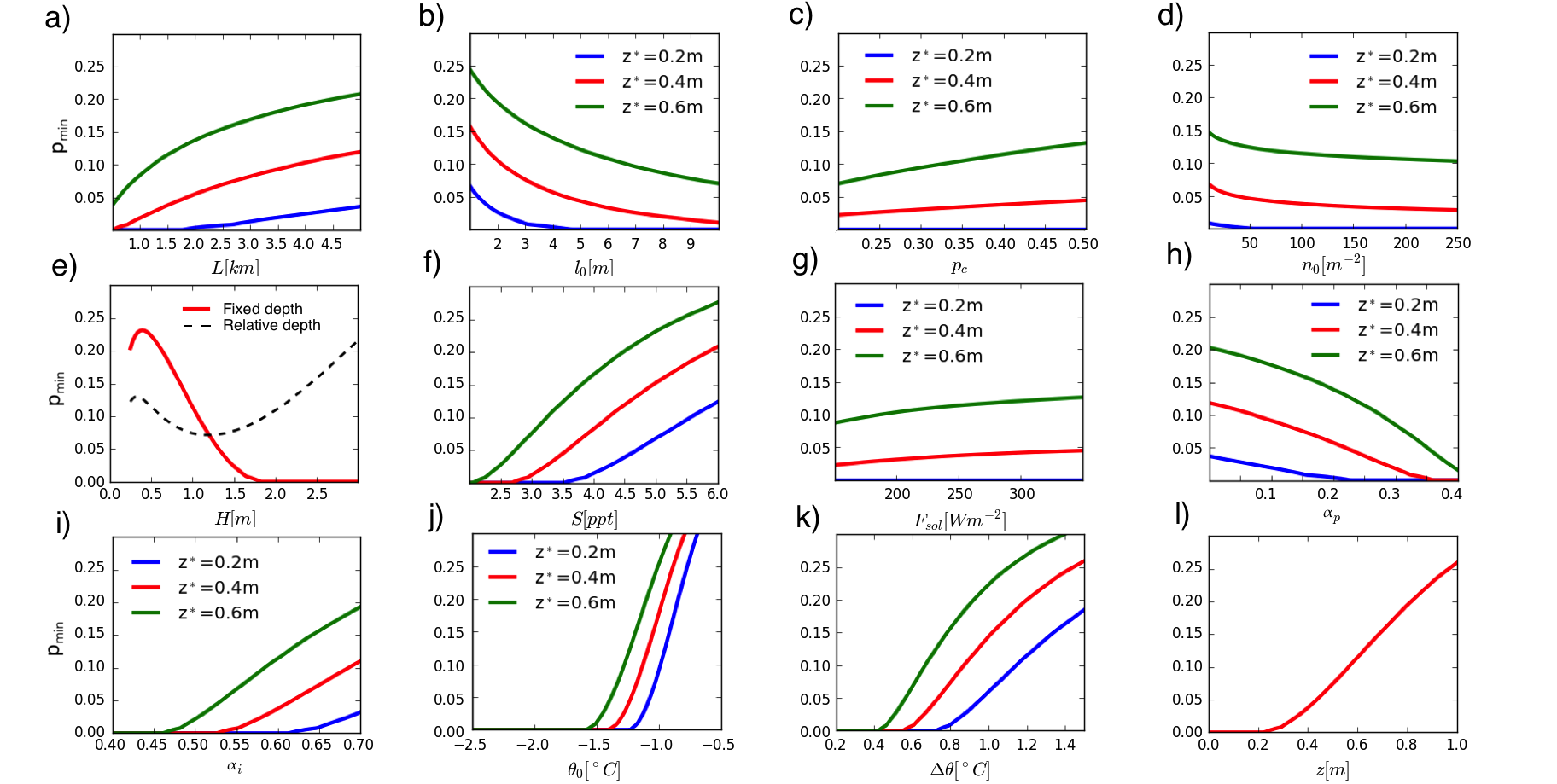}
\caption{Post-drainage pond coverage, $p_\text{min}$, as a function of physical parameters using Eqs. \ref{eqn:7}, \ref{eqn:9}-\ref{eqn:11}, and \ref{eqn:13} of the main text. In each panel, we are varying one parameter and assuming all other parameters are the defaults specified in Table \ref{tab:1} of the main text. We are assuming a cumulative normal hole opening distribution, $F$. Different colors in all panels except e and l stand for different values of the depth at which ice plugs tend to form, $z^*$. a) Pond coverage as a function of the basin length-scale, $L$. Here, we are keeping the density of brine channels, $n_0 = N_0/L^2$, constant. b) Pond coverage as a function of the typical pond length-scale, $l_0$. c) Pond coverage as a function of the percolation threshold, $p_c$. d) Pond coverage as a function of the brine channel density, $n_0 = N_0/L^2$. e) Pond coverage as a function of ice thickness, $H$. The red curve assumes that ice plugs form at a constant depth $z^*=0.6$ m. The black dashed curve assumes that ice plugs always form in the middle of the ice. f) Pond coverage as a function of ice salinity, $S$. g) Pond coverage as a function of solar flux, $F_\text{sol}$. h) Pond coverage as a function of pond albedo, $\alpha_p$. i) Pond coverage as a function of bare ice albedo, $\alpha_i$. j) Pond coverage as a function of the reference temperature, $\theta_0$. k) Pond coverage as a function of the temperature range, $\Delta \theta$. l) Pond coverage as a function of the depth where ice plugs form, $z^*$. }
\label{fig:params}
\end{figure*}

As a summary of our results in the main text, here we recapitulate how the minimum pond coverage, $p_\text{min}$, relates to physical parameters. After some manipulation of Eqs. \ref{eqn:5} to \ref{eqn:13} of the main text, we find
\begin{linenomath*}
\begin{gather}
p_\text{min} = p_c g\big( \eta_0 F(\tau_\text{m} - \tau_0 ) \big) \text{ , } \label{pdrepeat} \\[7pt]
\eta_0 \equiv c n_0 l_0^2 \quad \text{ , } \quad \tau_\text{m} \equiv \frac{T_\text{m}}{T_\text{h}} \quad \text{ , } \quad \tau_0 \equiv - F^{-1}\big(\frac{1}{n_0 L^2 }\big) \label{nondimpd} \\[11pt]
\tau_\text{m} \approx \underbrace{\frac{l_m}{(\alpha_i - \alpha_p) F_\text{sol}} \frac{\rho_w - \rho_i}{\rho_w} \frac{H}{1-p_\text{min}}}_{ T_\text{m} } \quad \underbrace{\frac{\theta_0^2 \Big( c^* k \frac{\theta_0}{H^2}  + (1-\alpha_p) F_\text{sol} \kappa e^{-\kappa z^*} \Big)}{\Delta \theta \rho_i \gamma S }}_{1/T_\text{h}} \label{taudyn}
\end{gather}
\end{linenomath*}
where $\alpha_i$ is the ice albedo, $n_0 \equiv \frac{N_0^2}{L^2}$ is the density of brine channels, and we have introduced the non-dimensional parameters $\eta_0$, the number of brine channels per characteristic area of the surface, $\tau_\text{m}$, the time for ponds to melt through the freeboard relative to the hole opening timescale, $T_\text{h}$, and $\tau_0$, the time between the first hole and the center of the hole distribution, $F$, relative to  $T_\text{h}$. Equation \ref{taudyn}, relates the non-dimensional timescale $\tau_\text{m}$ to measurable parameters using Eqs. \ref{eqn:5}, \ref{eqn:10} and \ref{eqn:11} of the main text. Note that $\tau_\text{m}$ depends on $p_\text{min}$ through the ``memorization'' timescale, $T_\text{m}$, so Eqs. \ref{pdrepeat} and \ref{taudyn} have to be solved simultaneously for $p_\text{min}$ and $\tau_\text{m}$.

The minimum pond coverage, $p_\text{min}$, is small when the differential melt is weak or when the drainage is efficient. A larger $\tau_\text{m}$ means that ponds melt quickly relative to the timescale of hole opening, so $p_\text{min}$ decreases with $\tau_\text{m}$. In particular, in terms of $T_\text{m}$ and $T_\text{h}$, when $T_\text{m}$ is large, melt is weak, so $p_\text{min}$ decreases with $T_\text{m}$, and when $T_\text{h}$ is large, holes open slowly, so $p_\text{min}$ increases with $T_\text{h}$. More brine channels per characteristic area increase the drainage efficiency, so $p_\text{min}$ decreases with $\eta_0$. A wider time separation between the initial hole and the center of the hole opening distribution means that ponds will spend more time in the tail of the distribution with few holes opening, decreasing the drainage efficiency, so $p_\text{min}$ increases with $\tau_0$. We now discuss the dependence of $p_\text{min}$ on each of the physical parameters that enter $\eta_0$, $\tau_\text{m}$, and $\tau_0$. We show these relationships in Fig. \ref{fig:params}, and we discuss each of the panels in the list below.  

\begin{enumerate}

\item In Fig. \ref{fig:params}a, we show that post-drainage pond coverage increases with size of the drainage basin, $L$. This is because a larger basin makes it more likely for the first hole to open somewhere within the domain, initiating the drainage stage earlier, and increasing $\tau_0$.

\item In Fig. \ref{fig:params}b, we show that minimum pond coverage decreases with the typical pond size, $l_0$. Physically, this is because larger ponds make it easier to drain the surface. Recall that it takes about one hole per pond of size $l_0$ to significantly drain the surface. Therefore, complete drainage can be achieved with fewer open holes if the individual ponds are larger.

\item Figure \ref{fig:params}c shows that minimum pond coverage increases with the percolation threshold, $p_c$. Pond coverage throughout its evolution from the end of the initial phase above the percolation threshold is approximately proportional to $p_c$. It is not exactly proportional since $\tau_\text{m}$ also depends on $p_c$, making the dependence of $p_\text{min}$ on $p_c$ slightly non-linear.

\item Figure \ref{fig:params}d shows the dependence of $p_\text{min}$ on the brine channel density, $n_0$. We can see that this dependence is relatively weak - the range of $n_0$ measured by \citet{golden2001brine}, from about $60 \text{ m}^{-2}$ to about $120\text{ m}^{-2}$, leads to only around 1\% change in pond coverage. Even an order of magnitude change in $n_0$ would lead to less than a 10\% change in pond coverage. Increasing $n_0$ leads to two competing effects. On the one hand, a higher $n_0$ increases $\eta_0$ by increasing number of potential holes, while on the other hand, it also increases $\tau_0$ by increasing the time before a significant fraction of holes open. These two effects largely cancel each other. As we show in section \ref{sec:holedist}, if the hole opening distribution, $F$, has an exponential tail, $F(x) \propto e^{-|x|}$, the two effects exactly cancel. Since in our case $F(x) \propto e^{-x^2}$ in the tail, some dependence on $n_0$ is retained, but the dependence remains weak. 

\item In Fig. \ref{fig:params}e, we show the dependence of $p_\text{min}$ on ice thickness, $H$. Ice thickness enters the non-dimensional parameter $\tau_\text{m}$ by increasing both the memorization and the hole opening timescale, $T_\text{m}$ and $T_\text{h}$. In particular, larger $H$ means that, on the one hand, ponds need to carve deeper depressions, thereby increasing $T_\text{m}$, while on the other hand, ice also warms more slowly, thereby increasing $T_\text{h}$. Because $H$ affects $T_\text{h}$, the effect of ice thickness depends on hole formation physics. In Fig. \ref{fig:params}e we show two scenarios. The red line shows a case where ice plugs form at a constant depth, $z^* = 0.6\text{ m}$. We can see that in this case pond coverage peaks at an ice thickness around 0.5 m after which it decreases to 0 with increasing thickness. The black dashed line shows a case where ice plugs always form in the middle of the ice. The dependence on ice thickness in this case is qualitatively different than the fixed depth scenario. After peaking at $H\approx 0.25\text{ m}$, pond coverage drops to its minimum at $H\approx 1.5\text{ m}$, after which it keeps increasing with increasing $H$. The complicated manner in which pond coverage depends on $H$ in both of these scenarios reflects two competing effects. So, depending on whether the effect on $T_\text{m}$ or $T_\text{h}$ is stronger, pond coverage either increases (if the effect on $T_\text{h}$ dominates) or decreases (if the effect on $T_\text{m}$ dominates) with $H$. We note that \citet{skyllingstad2015simulation} found that there was no clear relationship between ice thickness and post-drainage pond coverage in their detailed 3-dimensional model of ice and ponds.

\item In Fig. \ref{fig:params}f, we show that in our model $p_\text{min}$ increases with ice salinity, $S$. In our model, this is because saltier ice warms more slowly since a large fraction of energy is deposited in melting ice around brine pockets rather than going to warming the ice. This dependence is somewhat counterintuitive since a higher salt content means a larger brine volume fraction and therefore a larger potential for hole formation. For this reason it may be an artifact of our assumption that holes open when a certain temperature is reached. We note that \citet{skyllingstad2015simulation} did not find a significant relationship between salt content and minimum pond fraction in their model. 

\item Figure \ref{fig:params}g shows that $p_\text{min}$ depends only weakly on solar flux, $F_\text{sol}$. This weak dependence is explained by the two competing effects that nearly cancel each other. Namely, more solar radiation allows ponds to preferentially melt ice faster, decreasing $T_\text{m}$, while also warming the ice interior faster, thereby decreasing $T_\text{h}$.

\item In Fig. \ref{fig:params}h, we show how post-drainage pond coverage depends on pond albedo, $\alpha_p$. More reflective ponds prevent ice beneath ponds from warming quickly, increasing $T_\text{h}$, and also lead to a smaller contrast between bare ice and ponds, $\Delta \alpha$, increasing $T_\text{m}$ as well. However, the effect on $T_\text{m}$ dominates and pond coverage decreases with increasing $\alpha_p$.

\item In Fig. \ref{fig:params}i, we show how post-drainage pond coverage depends on ice albedo, $\alpha_i$. More reflective ice leads to a larger contrast between bare ice and ponds, decreasing $T_\text{m}$, which leads to a higher pond coverage.

\item  In Fig. \ref{fig:params}j, we show how $p_\text{min}$ depends on the reference temperature, $\theta_0$. This reference temperature is approximately the temperature at which holes tend to become open, and it is difficult to define more precisely without further investigation into the hole opening process. We can see that increasing $\theta_0$ leads to a rapid increase in pond coverage in our model. As we have discussed in section \ref{sec:heat}, the reference temperature affects the hole opening timescale by changing the warming rate. Namely, close to the melting point of the ice, heat capacity increases, rapidly decreasing the warming rate and increasing $T_\text{h}$ (see Fig. \ref{fig:heat_eq}c). Apart from increasing heat capacity, the warming rate also decreases since heat diffusion weakens when ice is warmer. As with salinity, this may be seen as counterintuitive since warmer ice contains a high brine volume fraction which would be expected to aid the formation of holes. 

\item In Fig. \ref{fig:params}k, we show that $p_\text{min}$ increases with the temperature range, $\Delta \theta$. This is due to the fact that it takes a longer time to warm a larger amount, increasing $T_\text{h}$. 

\item In Fig. \ref{fig:params}l, we show that $p_\text{min}$ increases with the depth at which ice plugs form, $z^*$. As less sunlight can penetrate deeper within the ice, ice warms slower, increasing $T_\text{h}$. A caveat here is that we have assumed that the rate of heat diffusion does not change with depth. As heat diffusion is likely strong near the bottom of the ice, this relation likely holds so long as the depth is not close to the ice bottom. 

\end{enumerate}
We believe that the assumptions we have made lead to dependencies on $L$, $l_0$, $p_c$, $n_0$, $F_\text{sol}$, $\alpha_p$, $\alpha_i$, $\Delta \theta$, and $z^*$ that are qualitatively correct. Despite this, it is difficult to make quantitative claims since some of these parameters, such as $\Delta \theta$ and $z^*$, are hard to constrain. In most panels of Fig. \ref{fig:params} we showed that pond coverage varies appreciably by changing the under-constrained parameter $z^*$, highlighting the quantitative uncertainty. We note that prior to this investigation, it was not recognized that geometric parameters $L$, $l_0$, and $p_c$ can have any effect on pond coverage. The dependence on parameters $H$, $S$, and $\theta_0$ seems problematic. Namely, as we have shown, the dependence on $H$ is highly sensitive to hole formation physics. The dependence on $S$ and $\theta_0$ is somewhat counter-intuitive, indicating that the assumption that temperature determines the onset of hole formation needs to be examined in more detail. The above analysis shows that, as we have already stressed several times, we need to improve our understanding of hole formation physics in order to understand how pond coverage depends on physical parameters.

Finally, we note that it is somewhat puzzling that the post-drainage coverage fraction, $p_\text{min}$, is neither 0 nor $p_c$ in a typical realistic situation. Namely, the function $g(\eta)$ behaves as $g(\eta) \rightarrow 0$ for $\eta \gg 1$ and $g(\eta) \rightarrow 1$ for $\eta \ll 1$, so there is only a limited range of $\eta$ for which $p_\text{min} \neq 0$ and $p_\text{min} \neq p_c$. The parameter $\eta$ falls within this range only when $T_\text{m} \sim T_\text{h}$, so a question then arises of why the time to melt through the thickness of the ice freeboard and the time for a significant number of holes to open are comparable. Partly, it must be because both phenomena are related to ice melt, but nevertheless, the two timescales fundamentally depend on different properties of the ice. Moreover, since there are many physical parameters that determine the value of $\eta$, its value should wildly fluctuate from one set of environmental parameters to another, making the outcome $\eta \ll 1$ or $\eta \gg 1$ likely. Therefore, the fact that we typically observe $p_\text{min} \neq 0$ and $p_\text{min}\neq p_c$ perhaps indicates that some of the physical parameters that control pond evolution are in fact not independent as we have assumed.



\newpage
\section{The hole opening distribution} \label{sec:holedist}

In the main text, we introduced a hole opening distribution, $F(\frac{t - t_0}{T_\text{h}})$, as a cumulative of some arbitrary probability density function, $f(\frac{t - t_0}{T_\text{h}})$. Here, we will describe some ways that this distribution affects pond evolution. 

First, we discuss how a time distribution, $f$, can be related to a measurable distribution of brine channel properties, $f_\theta$. In general, we may assume that there exists some underlying probability distribution, $f_\theta(\frac{\theta - \theta_0}{\Delta \theta})$, that describes the fraction of the holes that open when the ice interior temperature (or some other bulk ice property) increases from $\theta$ to $\theta + \text{d} \theta$. Such a distribution could in principle be measured in the field or modeled in some way. It is related to the time distribution, $f$, as 
\begin{linenomath*}
\begin{equation} \label{ftheta}
f(\frac{t - t_0}{T_\text{h}}) = f_\theta(\frac{\theta - \theta_0}{\Delta \theta}) \frac{\text{d} \theta}{\text{d} t} \text{  . }
\end{equation}
\end{linenomath*}
This equation is valid even when the ice warming rate is not constant and provides a basis to relate the hole opening distribution that enters Eqs. \ref{eqn:12} and \ref{eqn:14} of the main text to measurable sea ice properties.

Next, we discuss the variability of the distribution center, $t_0$. In the main text, we remarked that the center of the hole opening distribution, $t_0$, is approximately (see Eq. \ref{eqn:9} of the main text)
\begin{linenomath*}
\begin{equation} \label{approxt0}
t_0 \approx -T_\text{h} F^{-1} \big(\frac{1}{N_0} \big) \text{  . }
\end{equation}
\end{linenomath*}
We argued that, because the timing of the opening of the first hole fluctuates, the time until a given fraction of holes, $F(0)$, opens also fluctuates, and, consequently, so does $t_0$. Here we will discuss the probability distribution of $t_0$ over an ensemble of runs. Let us assume that the first hole opens at some temperature $\theta_i$. $t_0$ is then the time it takes for ice to warm from $\theta_i$ to $\theta_0$. There are $N_0$ holes and each one opens at a different critical temperature independently drawn from a distribution $f_\theta$. Thus, $\theta_i$ is a minimum of a set of $N_0$ independent random variables drawn from a distribution $f_\theta$. Therefore, the probability of finding some $\theta_i$, and a corresponding $t_0$, is governed by extreme value statistics (see, e.g., \citet{coles2001introduction} for an introduction to extreme value statistics). 

In general, depending on the tail of an underlying distribution, we expect the probability of finding some minimum value from a large sample of independent random variables to fall into one of three universal distributions. In our case, we expect that $f_\theta$ has a relatively well-defined width. A distribution with a well-defined width will likely have an exponential tail, and such distributions have extreme value statistics described by a Gumbel distribution
\begin{linenomath*}
\begin{equation}
f_\text{Gumbel}(x) = \frac{1}{\beta} e^{-(\frac{x-\mu}{\beta} + e^{-\frac{x-\mu}{\beta}})} \text{  , }
\end{equation}
\end{linenomath*}
where $\mu$ is the mode of the distribution and $\beta$ is the scale. Therefore, we can expect that, for each run, $t_0$ is drawn from this distribution with a mode that is close to our approximation in Eq. \ref{approxt0}. Exactly how parameters $\mu$ and $\beta$ depend on $N_0$ and $T_\text{h}$ depends on the choice of $F$. Note that this distribution determines the probability of finding $t_0$ on an ensemble of runs, in contrast with the distributions $f$ and $f_\theta$ that determine the probability of opening a brine channel within a single run. Because of this intrinsic variability, even runs with identical bulk parameters can end up with different pond coverage. In our simulations this variability accounted for about $5\%$ of the variability in the minimum pond coverage, $p_\text{min}$. 

Finally, we show that it is mainly the tail of the hole opening distribution, $F$, that controls pond evolution. We do not know $F$ for real ice, and it can in principle have any shape as long as it is a function monotonically increasing from 0 to 1. The universal function, $g(\eta)$, behaves as $g(\eta) \rightarrow 1$ for $\eta \rightarrow 0$ and $g(\eta) \rightarrow 0$ for $\eta \rightarrow \infty$. In particular, $g$ falls below $0.1$ for $\eta \sim 10$. Therefore, when $\eta$ exceeds a value of around 10, $g$ falls to approximately zero, and further increase in $\eta$ does not affect pond evolution. Recall that $\eta  = \eta_0 F(\frac{t - t_0}{T_\text{h}})$, where $\eta_0 \equiv c\frac{l_0^2}{L^2}N_0$. If, we assume reasonable parameter values $c \sim 3$, $l_0 \sim 5\text{ m}$ \citep{popovic2020snow}, and $\frac{N_0}{L^2} \sim 100 \text{ m}^{-2}$ \citep{golden2001brine}, we get an order of magnitude estimate $\eta_0 \sim 10000$. Therefore, $\eta \sim 10$ when $F\sim 0.001$ and pond evolution only occurs in the tail of the distribution $F$. Thus, due to the large density of brine channels, the exact shape of $F$ does not matter apart from determining the weight of the distribution that falls in the tail. We discuss this in more detail below. 

Let us look at some particular examples of $F$ to see exactly how it affects pond evolution. As a first example, let us assume that the probability density function, $f$, falls off exponentially, $f(x) \propto e^{x}$ for $x \ll -1$. In this case, $F(x) = Ae^{x}$ for $x \ll -1$ and $F^{-1}(y) = \ln{y/A}$ for $0 < y \ll 1$, where $A$ is a normalizing constant that also takes into account the rest of the distribution that does not fall in the tail. If we use the approximate relationship $t_0 \approx -T_\text{h}F^{-1}(1/N_0)$ (see Eq. \ref{approxt0}), we find 
\begin{linenomath*}
\begin{equation}
\eta(t) \approx c\frac{l_0^2}{L^2}N_0 A e^{t/T_\text{h} - \ln{AN_0}} =  c\frac{l_0^2}{L^2} e^{t/T_\text{h} } \text{  , } 
\end{equation}
\end{linenomath*}
for $t \ll T_\text{h} \ln{AN_0}$. Therefore, if the tail of the hole opening distribution is exponential, pond evolution is independent of the brine channel density, $N_0/L^2$. Pond evolution is also independent of the shape of the distribution beyond the tail as seen from the fact that the normalizing constant, $A$, does not appear in the equation for $\eta$. If we recall that pond coverage during stage II is $p = p_c g(\eta(t))$ and note that $g$ is approximately logarithmic in $\eta$ within the region where pond coverage varies from $p_c$ to 0, we see that pond coverage falls approximately linearly with time. 

As a second example, we consider $f$ consistent with a normal distribution, $f(x) \propto e^{-x^2}$, which we used in the main text to compare to observations. In this case, $F(x) = Ae^{-x^2}$ for $x \ll -1$ and $F^{-1}(y) = -\sqrt{\ln{A/y}}$ for $0 < y \ll 1$, with $A$ being a normalizing constant that accounts for the distribution beyond the tail. Following similar logic as above, we can derive
\begin{linenomath*}
\begin{equation}
\eta(t) \approx  c\frac{l_0^2}{L^2} e^{- (t/T_\text{h})^2 + 2t\sqrt{\ln{AN_0}}/T_\text{h}} \text{  . }
\end{equation}
\end{linenomath*}
Here, we see that due to the term $\sqrt{\ln{AN_0}}$, dependence on $N_0$ and $A$ is not lost. We note two facts about this dependence - 1) pond evolution depends not on the density of brine channels but explicitly on the \textit{number} of brine channels within a drainage basin, and 2) the shape of the distribution beyond the tail only acts to modify the effective number of brine channels through $AN_0$. This unexpected dependence on the total number of brine channels is a consequence of the fact that stage II begins when the first hole opens rather than when a certain fraction of holes open. We discussed this in section \ref{sec:params}. We add that the dependence on $N_0$ and $A$ is relatively weak - a 10-fold change in the effective number of brine channels leads to only around a $10\%$ change in $\sqrt{\ln{AN_0}}$, and a correspondingly small change in the rate of change of pond coverage. As noted in \citet{golden2001brine}, typical brine channel densities are between $60\text{ m}^{-2}$ and $120\text{ m}^{-2}$, so the uncertainty in the number of channels is likely not a significant source of error so long as we can reliably estimate the size of the drainage basin, $L$. We also believe that the discussion above partially explains why our rather ad-hoc choice of $F$ as a cumulative normal distribution gave good agreement with observations. Namely, even if the majority of this distribution is not normal, the prediction about pond evolution will be accurate so long as the actual hole opening distribution falls off as $e^{-x^2}$.

For the more general case of $F = A e^{-|x|^a}$ in the tail, we can follow similar logic to find that as the power $a$ increases, the dependence on $N_0$ and $A$ becomes stronger. Pond coverage depends most strongly on $N_0$ when the distribution $F$ has a sharp cutoff.

\newpage

\bibliographystyle{plainnat}
\bibliography{mybibliogJGR.bib}

\makeatletter\@input{xx_ms_percolation.tex}\makeatother